\documentclass[prd,aps,superscriptaddress,nofootinbib]{revtex4}
\usepackage[latin9]{inputenc}
\usepackage{amsmath}
\usepackage{amssymb}
\usepackage{graphicx}

\makeatletter

\providecommand{\tabularnewline}{\\}

\@ifundefined{textcolor}{}
{%
 \definecolor{BLACK}{gray}{0}
 \definecolor{WHITE}{gray}{1}
 \definecolor{RED}{rgb}{1,0,0}
 \definecolor{GREEN}{rgb}{0,1,0}
 \definecolor{BLUE}{rgb}{0,0,1}
 \definecolor{CYAN}{cmyk}{1,0,0,0}
 \definecolor{MAGENTA}{cmyk}{0,1,0,0}
 \definecolor{YELLOW}{cmyk}{0,0,1,0}
}

\usepackage{multirow}\usepackage{float}\@ifundefined{definecolor}
 {\usepackage{color}}{}

\makeatother

\begin{document}
\begin{flushright}
Preprint SMU-HEP-13-09 
\par\end{flushright}

\vspace{0.2in}

\title{Charm quark mass dependence in a global QCD analysis}

\author{Jun Gao}

\affiliation{Department of Physics, Southern Methodist University, Dallas, TX
75275-0181, USA}

\author{Marco Guzzi}

\affiliation{Deutsches Elektronen Synchrotron DESY, Notkestrasse 85 D-22607 Hamburg,
Germany}

\author{Pavel Nadolsky}

\affiliation{Department of Physics, Southern Methodist University, Dallas, TX
75275-0181, USA}
\begin{abstract}
We study the effect of the charm quark mass in the CTEQ global analysis
of parton distribution functions (PDFs) of the proton. Constraints
on the $\overline{\rm MS}$ mass of the charm quark are examined at the
next-to-next-to-leading order (NNLO) accuracy in the S-ACOT-$\chi$
heavy-quark factorization scheme. The value of the charm quark mass
from the hadronic scattering data in the CT10 NNLO fit, including
semiinclusive charm production in DIS at HERA collider, is found
to agree with the world average value. Various approaches for constraining
$m_{c}$ in the global analysis and impact on LHC cross sections are
reviewed. \\

\noindent \textbf{Keywords}: Global Analysis, Charm Quark, Parton
Distributions 

\noindent \textbf{PACS}: 14.65.Dw, 12.38.-t 
\end{abstract}
\maketitle

\section{Introduction}

Quark masses are free parameters of the QCD Lagrangian that parametrize
explicit breaking of the chiral symmetry. For quarks heavier than
1 GeV, quark masses arise as independent hard scales $m_{Q}$ in perturbative
QCD calculations for particle cross sections. Quarks are not observed
freely because of color confinement, hence their properties and masses
are established indirectly by comparing theory calculations against
experimental data on hadronic reactions. In a global analysis of parton
distribution functions (PDFs) of the proton, the method by which the
heavy-quark masses are included in experiments at energies comparable
to $m_{Q}$ has non-negligible impact on the extracted PDFs \cite{Tung:2006tb}.
Variations in the PDFs associated with the treatment of heavy quarks
have phenomenological consequences for electroweak precision measurements
at the Large Hadron Collider \cite{Nadolsky:2008zw}.

In this paper we explore constraints on the charm quark mass from
the global hadronic data in the CTEQ NNLO PDF analysis. The study
is motivated by publications of combined cross sections on inclusive
deep-inclusive scattering (DIS) 
and semi-inclusive DIS charm production at the $ep$ collider
HERA \cite{Aaron:2009aa,Abramowicz:1900rp}. Among all experimental data
sets included in the global fit, the DIS experiments have
the best potential to constrain the charm mass. 
On the theory side, PQCD calculations
for neutral-current deep inelastic scattering 
have been extended to the 2-loop level in the QCD coupling strength
$\alpha_{s}$ both for massless
\cite{SanchezGuillen:1990iq,vanNeerven:1991nn,Zijlstra:1991qc} and massive 
\cite{Laenen:1992zk,Riemersma:1994hv,Harris:1995tu} quarks, 
while massless \cite{Moch:2004xu,Vermaseren:2005qc}
and some massive \cite{Blumlein:2006mh,Bierenbaum:2009mv,Ablinger:2010ty,Blumlein:2012vq,Ablinger:2011pb,Ablinger:2012qj,Ablinger:2012qm,Ablinger:2012sm}
coefficient functions were also obtained at the 3-loop level. 
With such accuracy, it becomes possible to determine the charm quark
mass and its uncertainty from the DIS data.

The current world average of the charm mass 
in the $\overline{\textrm{MS}}$ renormalization scheme 
is $m_{c}(m_{c})=1.275\pm 0.025$ GeV~\cite{Beringer:1900zz}.
This value is derived primarily from
measurements in timelike scattering processes 
and lattice simulations, using analyses that include 
up to four orders in perturbative QCD. 
The precise DIS data from HERA in principle 
allows us to extract $m_{c}(m_{c})$ from a spacelike
scattering process and compare to other determinations.

In many previous PDF analyses, the heavy-quark masses have been treated
as effective parameters rather than fundamental constants. They were
anticipated to deviate from the $\overline{\rm MS}$ masses or even be
fully independent. Besides entering the exact matrix elements,
the heavy-quark masses control various approximations in DIS calculations, 
as will be reviewed below. The approximations 
affect extraction of the mass values from the hadronic data, but
their effect is of a higher order in the QCD coupling
strength according to the QCD factorization theorem. Thus, although
the bias due to the approximations can be important at low orders, it
is expected to subside as more 
loops are included in PQCD calculations. At high enough order of $\alpha_s$, 
such as the NNLO, a comparison to the world-average $\overline{\rm MS}$ quark
mass becomes feasible. 

There exist several theoretical approaches, or heavy-quark schemes,
for computations involving massive quarks \cite{Aivazis:1993pi,Buza:1996wv,Chuvakin:1999nx,Thorne:1997uu,Thorne:1997ga,Thorne:2006qt,Forte:2010ta}. 
The extracted $m_{c}$ value depends on the heavy-quark
scheme as well as the order of the PQCD calculation. In this work,
we adopt a two-loop implementation \cite{Guzzi:2011ew}
of the general-mass 
scheme S-ACOT-$\chi$ \cite{Aivazis:1993pi,Kramer:2000hn,Tung:2001mv}
employed in the CT10 NNLO analysis \cite{Gao:2013xoa}. We discuss physics
assumptions affecting the extracted value 
of $m_c(m_c)$ and comparisons to other
recent extractions of $m_{c}$ \cite{Martin:2010db,Abramowicz:1900rp,Alekhin:2012vu}.
Within this scheme, we find the range of input $m_{c}(m_{c})$ values
providing the best description of the CT10 fitted data and compare it 
to the world-average value. Finally, we analyze the impact 
of the uncertainty in the charm mass on benchmark LHC predictions.

\section{Charm mass definitions and NNLO predictions for DIS\label{sec:Charm-mass-definitions}}

\subsection{Overview\label{sec:TheoryOverview}}

Quantum field theory operates with two common definitions of the quark
mass, the pole mass and the $\overline{\textrm{MS}}$ mass. The pole
mass is defined as the position of the pole in the renormalized quark
propagator. The pole mass is infrared-safe, gauge-invariant, and is
derived in the on-shell renormalization scheme. It is often close
to the experimental mass definition 
\cite{Gray:1990yh,Chetyrkin:1999qi,Melnikov:2000qh,Marquard:2007uj},
but, as the pole charm mass value of 1.3-1.8 GeV 
borders the nonperturbative region, 
accuracy of its determination is limited by significant radiative
contributions associated with renormalons
\cite{Bigi:1994em,Beneke:1994sw,Beneke:1998ui}. Because of 
large perturbative coefficients arising even at three or
four loops in the QCD coupling $\alpha_{s}$, the pole $m_{c}$ value
cannot be determined to better than a few hundred MeV. 

The $\overline{\textrm{MS}}$ mass $m_{c}(\mu)$, on the other hand,
is the renormalized quark mass in the modified-minimal-subtraction
scheme, defined as a short-distance mass that is not affected by nonperturbative
ambiguities. It is evaluated at a momentum scale $\mu$ typical for
the hard process, frequently taken to be the mass $m_{c}$ itself. 
Precise determinations of $m_{c}(m_{c})$ achieve 
a smaller uncertainty of order 30 MeV or
less. The $\overline{\textrm{MS}}$ mass starts to differ from the
pole mass beginning at order ${\cal O}(\alpha_{s})$. The conversion
between the $\overline{\textrm{MS}}$ mass to the pole mass is required
in the PDF analysis, as the massive 2-loop Wilson
coefficients and operator matrix elements in DIS are available in
terms of the pole mass. The conversion procedure will be reviewed 
in the next section.  

In parton-level diagrams for deep-inelastic scattering, external massive
quarks may arise both as the final and initial states. For the
quarks that are heavier than the proton, some factorization schemes 
introduce an effective PDF to describe their quasi-collinear
production at high energy. The heavy-quark PDF can contribute 
to the hadronic cross section
through a convolution with a hard-scattering matrix element with the
heavy quark(s) in the initial state, also called a ``flavor-excitation''
matrix element. In contrast, the ``flavor-creation'' matrix elements include
only light quarks and gluons in the initial state, while the heavy
quarks are only in the final state. The ``flavor-excitation''
contributions commonly arise in the variable flavor number (VFN)
schemes, such as the general-mass VFN (GM VFN) scheme. The alternative
fixed-flavor number (FFN) scheme does not introduce a heavy-quark PDF 
operates with ``flavor-creation'' terms only. 

For this study we employ the S-ACOT-$\chi$ general-mass scheme \cite{Aivazis:1993pi,Kramer:2000hn,Nadolsky:2009ge}
implemented to the 2-loop (NNLO) accuracy \cite{Guzzi:2011ew}.
This scheme includes exact massive flavor-creation contributions that
dominate at low boson virtualities $Q$, as well as
the approximate flavor-excitation
terms that are important at high $Q$. Thus, the S-ACOT-$\chi$ scheme
reduces to the FFN scheme at $Q^{2}\approx m_{c}^{2}$
and to the zero-mass VFN scheme at $Q^{2}\gg m_{c}^{2}$. 

In a  comprehensive factorization scheme such as GM-VFN, the charm mass plays several roles. First, the \emph{exact}
charm mass enters Feynman diagrams for charm
particle creation in the final state, such as $\gamma^{*}g\rightarrow c\bar{c}$
in NC DIS. Second, auxiliary scales are introduced that
are of order of the fundamental charm mass (either the pole mass 
or $\overline{\rm MS}$ mass), but need not to coincide with it. 

One such scale sets the energy for switching from the 3-flavor
to 4-flavor evolution in the running $\alpha_{s}(\mu)$,
which is utilized by both FFN and VFN computations.
A similar switching scale from 3-flavor evolution to 4-flavor evolution
arises in the PDFs $f_{a/p}(x,\mu)$.
The charm mass also defines characteristic energy scales in the 
flavor-excitation contributions, cf. Sec.~\ref{sec:Details-of-implementation}.
Finally, there can be auxiliary scales associated with the final-state
quark fragmentation into hadrons, present both in the FFN and GM-VFN schemes. 
The dependence on these auxiliary scales is reduced
with each successive order of perturbation theory.

When the input mass is varied in the global fit, the response of the
DIS cross sections reflects coordinated variations of all such scales.
An important question arises when interpreting the outcome of the
fit: which  mass parameter controls the agreement with the data, the 
exact charm mass or the approximate mass in the auxiliary scales?

We have found that the global fit is sensitive to the exact charm mass, 
despite the introduced approximations. 
In one exercise, we have
varied the input charm mass in the exact DIS coefficient functions 
for flavor-creation processes, while keeping it fixed in the above
auxiliary mass scales.
In a complementary exercise, we varied the input mass in all auxiliary
scales, while keeping it fixed in the exact flavor-creation
coefficient functions. In both cases, we examined the agreement 
with the data as a function of the varied mass parameter.
We followed the fitting procedure outlined in the next section and
assumed fixed PDF parametrizations for the best-fit $m_c(m_c)$ found in
the main analysis.

\begin{figure}[h!]
\begin{centering} 
\includegraphics[width=0.48\textwidth]{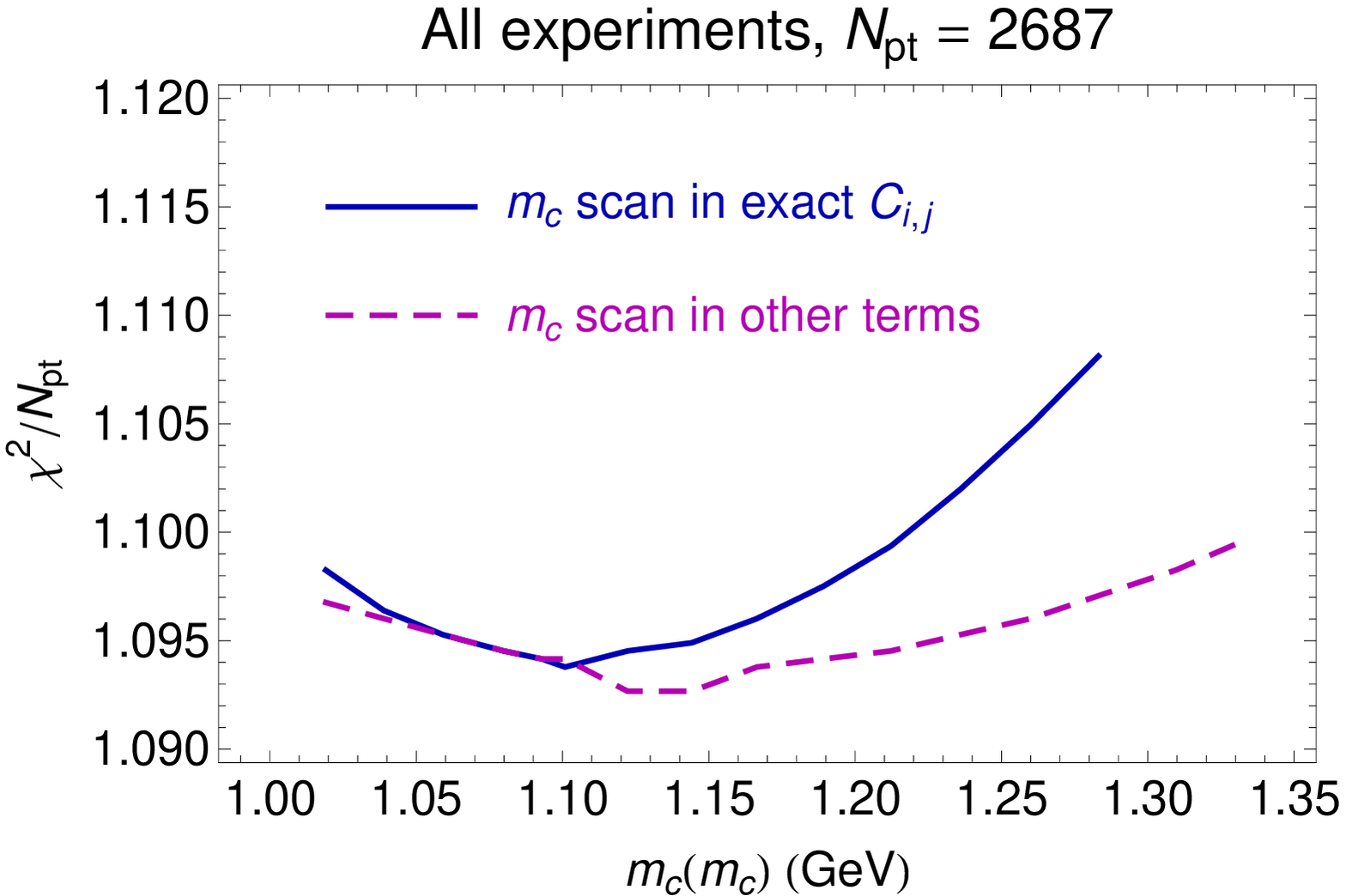}~\includegraphics[width=0.48\textwidth]{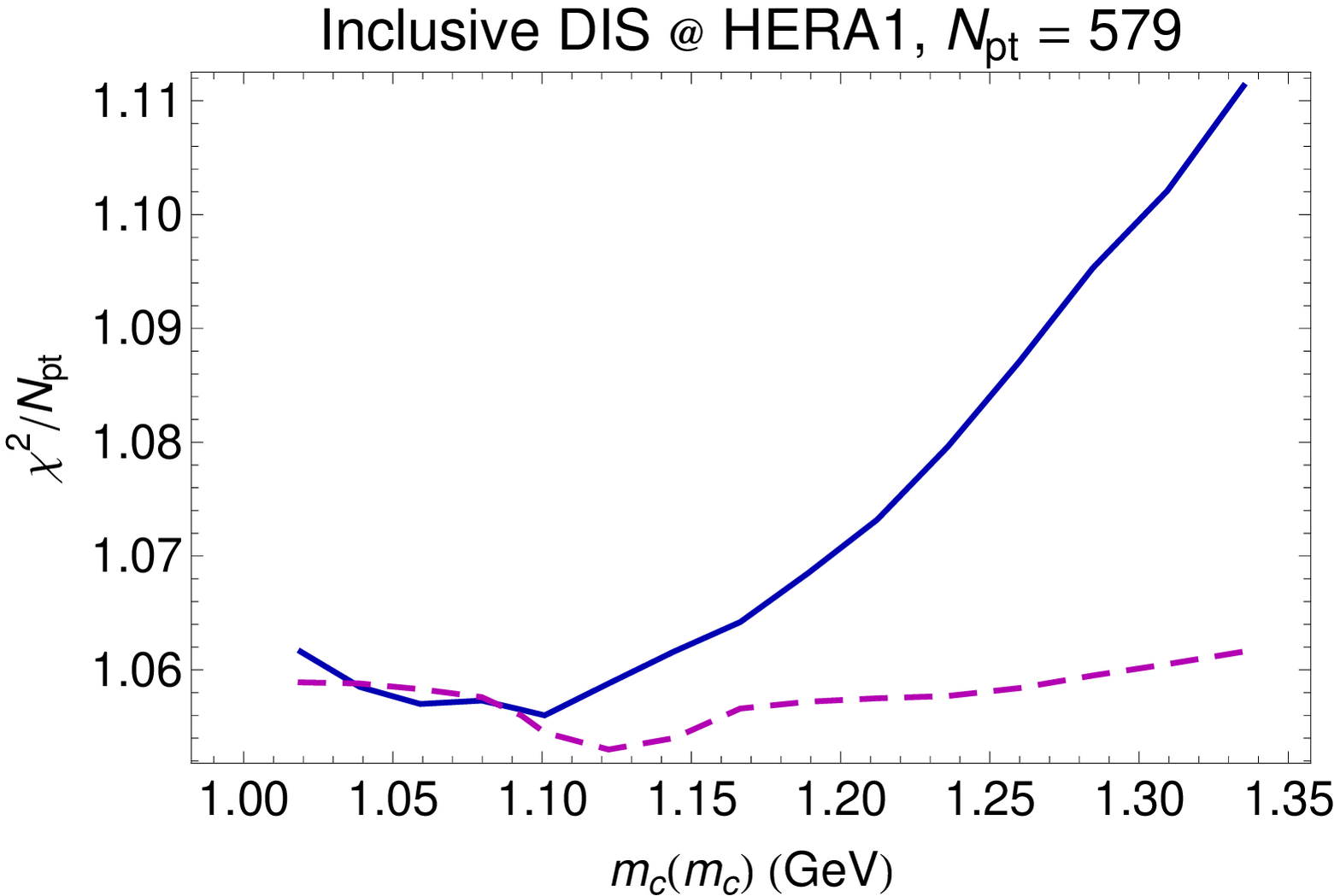}\\
\includegraphics[width=0.48\textwidth]{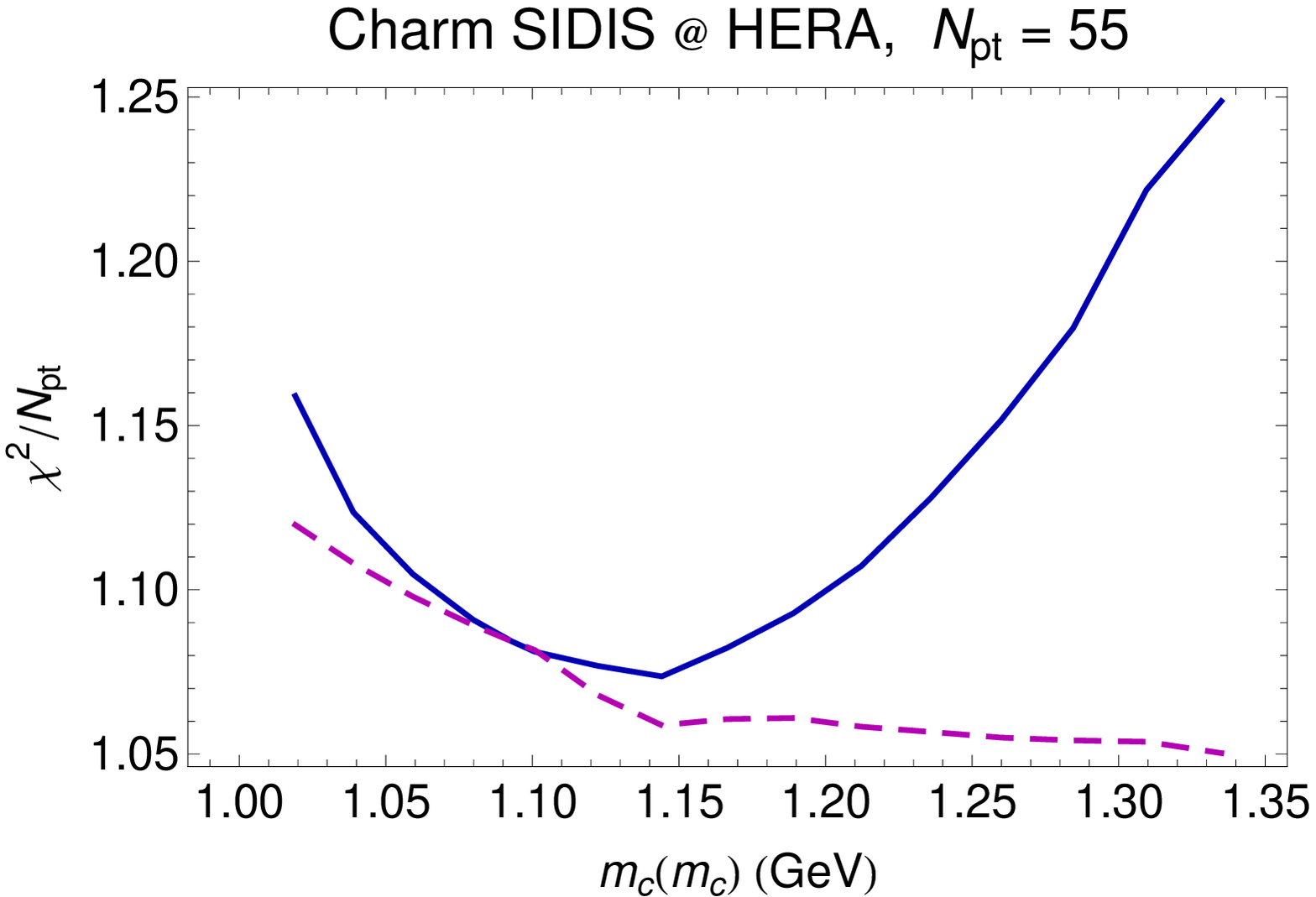}
\par\end{centering}

\caption{\label{fig:chi2separate} Dependence of $\chi^{2}/N_{pt}$ as a function
of $m_{c}(m_{c})$ in the exact flavor-creation coefficient functions
(solid blue lines) and auxiliary energy scales listed in the text
(dashed magenta lines). }
\end{figure}

The dependence of the figure-of-merit function $\chi^{2}$ on the
varied $m_{c}(m_{c})$ in these exercises is shown 
for all fitted experiments, 
the combined HERA inclusive DIS data \cite{Aaron:2009aa},
and the combined HERA semi-inclusive charm production data \cite{Abramowicz:1900rp}
in the upper left, upper right, and lower panels of Fig.~\ref{fig:chi2separate}.
To better visualize the comparison, $\chi^2$ is divided by the
number $N_{pt}$ of data points for each data set. 
The solid blue line and dashed magenta line are for $\chi^2/N_{pt}$
for the varied mass parameter in the exact DIS
coefficient functions and in the auxiliary mass scales, respectively.

With all PDF parameters fixed, variations in $\chi^{2}$ are more
pronounced in the mass scan of the first type, when the charm mass
is varied only in the exact DIS coefficient functions. In this case,
$\chi^2$ in both inclusive and semiinclusive DIS shows a pronounced
minimum as a function of $m_c(m_c)$. 

In the second case, when $m_{c}(m_{c})$
is varied only in the auxiliary scales, the $\chi^{2}$ dependence 
is flatter and has a shallow minimum at most. 
This exercise indicates that both inclusive and semiinclusive DIS
cross sections are more sensitive to the exact $m_{c}$ mass 
in the flavor-creation coefficient functions than to the auxiliary
scales. The detailed constraints on $m_{c}$ are determined by the
interplay of these two trends as well as by variations 
in the PDFs and other inputs. 

\subsection{Details of implementation \label{sec:Details-of-implementation}}

\subsubsection{Conversion to the pole mass}
Our calculation proceeds by taking the $\overline{\rm MS}$ quark masses
as the input for the whole calculation. The transition from the 3-flavor
to 4-flavor evolution in $\alpha_{s}$ and PDFs is taken to occur
at the scale equal to this input mass.%
\footnote{The evolution of $\alpha_{s}$ and PDFs is carried out using the HOPPET
computer code \cite{Salam:2008qg}, configured so that transitions from 
$N_{f}$ to $N_{f+1}$ flavors occur at the $\overline{\rm MS}$ masses.%
}

The massive 2-loop contributions to neutral-current DIS include matrix 
elements with explicit creation of $c\bar{c}$ pairs \cite{Riemersma:1994hv} 
and operator matrix elements $A_{ab}^{(k)}$ \cite{Buza:1996wv} for collinear production of massive quarks. Their available expressions take the pole mass as the input. For these parts, the $\overline{\rm MS}$
mass is converted to the pole mass according to the perturbative
relation in Eq. (17) of \cite{Chetyrkin:2000yt},
\begin{eqnarray}
m^{pole}_Q&=&m_Q(m_Q)\ \Biggl\{ 1 +
\frac{\alpha_s(m_Q(m_Q),N_f)}{\pi}\frac{4}{3} \nonumber\\
&+& 
\frac{\alpha_s^2 (m_Q(m_Q),N_f)}{\pi^2}\Bigl[13.4434-1.04137 N_f+ 
\frac{4}{3}\sum_{i=1}^{N_f}\Delta(m_i(m_i)/m_Q(m_Q))\Bigr] \Biggr\},
\label{msbar2pole}
\end{eqnarray}
where $\Delta (x)= 1.2337\ x - 0.597\ x^2 +0.23 \ x^3$, 
and the coefficients are shown up to order $\alpha_s^2$. 

For the charm quark, the $\alpha_s$ and $\alpha_s^2$ contributions 
in the conversion formula have comparable magnitudes,\footnote{For
  example, $m_c(m_c)=1.15$ GeV translates into 
$m_c^{pole}=1.31, 1.54, 1.86$ GeV using one, two, three loops in the
  conversion formula with $\alpha_s(M_Z)=0.118$.} the procedure
for $m_c^{pole}$ truncation is numerically important. We implement two
conversion methods. In one method, the $\overline{\rm
  MS}$ mass is always converted to the pole mass via the full 2-loop 
relation. Alternatively, the 2-loop (1-loop) conversion is performed 
within the ${\cal O}(\alpha_s)$ and ${\cal O}(\alpha_s^2)$ terms of the 
Wilson coefficient functions and OME's, respectively. This is argued
to be equivalent to calculating DIS structure functions directly  
in terms of the $\overline{{\rm MS}}$ mass and improve
perturbative convergence of the $m_{c}(m_c)$ fit~\cite{Alekhin:2010sv}. Numerically, it is not obvious yet that the second (truncated) method
improves perturbative convergence at the implemented orders: 
the effect of including an approximate 3-loop correction in the second method 
\cite{Alekhin:2012vu} (producing the change 
$\delta m_c(m_c)\approx 0.09 $ GeV compared to
the 2-loop result) exceeds  
the difference between two conversion methods at two loops 
($\delta m_c(m_c)\approx 0.07 $ GeV).

In charged-current DIS,
for which only 1-loop expressions are included, we can use either
the $\overline{\rm MS}$ mass or pole mass and find almost no sensitivity
to the choice. 

\subsubsection{Rescaling variable}
The general-mass schemes \cite{Buza:1996wv,Chuvakin:1999nx,Thorne:1997uu,Thorne:1997ga,Thorne:2006qt,Forte:2010ta}
differ primarily in the form of approximation for flavor-excitation
coefficient functions (with incoming heavy quarks) at $Q$ comparable
to $m_{c}$. When $m_{c}$ is negligible, the coefficient functions
in all GM-VFN schemes reduce to unique zero-mass expressions,
but, near the threshold, they may differ by powerlike contributions
$(m_{c}^{2}/Q^{2})^{p}$ with $p>0$ in the approximate flavor-excitation
terms. The powerlike corrections are suppressed in the full result 
by an additional power of
$\alpha_{s}$, \emph{i.e.}, they are of order $\alpha_{s}^{3}$ in
an NNLO ($\alpha_{s}^{2}$) calculation.

In the S-ACOT-$\chi$ scheme, the form of the powerlike contributions
is selected based on the general consideration of energy-momentum
conservation  \cite{Tung:2001mv,Tung:2006tb,Nadolsky:2009ge}. 
As a result, the flavor-excitation contributions 
are suppressed at energies close to the massive quark production threshold,
producing better description of the DIS data. 
The flavor-excitation coefficient functions are constructed 
from the respective zero-mass coefficient functions $c_{ZM}(\chi/\xi,Q/\mu)$, where 
the parton's momentum fraction $\chi$ is rescaled with respect to Bjorken $x$ by a factor dependent on the total mass $M_f$ of heavy quarks in the final state. The general behavior of $\chi$ is determined from the
condition of the threshold suppression, while its detailed form can
be varied to estimate the associated higher-order 
uncertainty in the extracted $m_{c}$. 
This approach is readily demonstrated \cite{Guzzi:2011ew} to be compliant 
with the QCD factorization theorem for DIS cross sections to all orders 
\cite{Collins:1998rz}, which is one of the advantages 
of using the S-ACOT-$\chi$ scheme.

In the default implementation, the momentum fraction
in NC DIS charm production is given by $\chi=x(1+4m_{c}^{2}/Q^{2})$, corresponding
to $M_{f}=2m_{c}$ for the lightest final state ($c\bar{c}$).
\footnote{Starting from $O$($\alpha_{s}^{2}$), contributions with up to four
massive quarks in the final state can appear. In such terms, we still 
use $\chi=x\left(1+4m_{c}^{2}/Q^{2}\right)$, given
their smallness in the total result \cite{Guzzi:2011ew}.
} In charged-current DIS, we set
$\chi=x\left(1+m_{c}^{2}/Q^{2}\right)$. The rescaling 
ratio $\chi/x$ is thus independent of $x$. On 
general grounds, it may be expected that the threshold suppression
is less pronounced at $W^{2}=Q^{2}(1/x-1)\rightarrow\infty$ for fixed
$Q$, corresponding to $x\rightarrow0$. In this limit, it may be
desirable to reduce or even eliminate the rescaling altogether, 
as quasi-collinear production of heavy quarks becomes
more feasible. 

To allow for this possibility, a generalized rescaling variable $\zeta$
can be implicitly defined by~\cite{Nadolsky:2009ge}
\[
x=\zeta\, \left(1+\zeta^{\lambda}M_{f}^{2}/Q^{2}\right)^{-1},
\]
where $\lambda$ is a positive parameter, typically $0\leq
\lambda \lesssim 1$. The S-ACOT-$\chi$ scheme
is reproduced with $\lambda=0$, and the rescaling is fully turned off for $\lambda\gg 1$. For $\lambda\neq0$,
the mass-threshold constraints are enforced at small $W$ (large $x$),
but the standard $x$ variable is recovered at large $W$ 
(small $x$) in a controllable way.

The sensitivity of the CTEQ global fit to $\lambda$ is of the same
order as the sensitivity to the $m_c^{pole}$ truncation method. 
The changes in the preferred value of $m_{c}$ that we observe provide
an estimate of the uncertainty due to the powerlike corrections.

\section{Charm mass scans \label{sec:Results}}
\subsection{Setup}
Using the theoretical setup reviewed in the previous section and the general procedure of the  CT10 NNLO PDF analysis \cite{Gao:2013xoa}, a scan of
the log-likelihood function $\chi^{2}$ over the input $\overline{\rm MS}$
charm mass was carried out in the range $1\leq m_{c}(m_{c})\leq1.4$ GeV.
Besides the combined HERA data on inclusive DIS and semiinclusive
charm production, we include experimental data from DIS measurements
by BCDMS~\cite{Benvenuti:1989rh,Benvenuti:1989fm}, NMC~\cite{Arneodo:1996qe},
CDHSW~\cite{Berge:1989hr}, and CCFR~\cite{Yang:2000ju,Seligman:1997mc};
NuTeV and CCFR dimuon production~\cite{Goncharov:2001qe,Mason:2006qa};
$F_{2c}$ measurements at HERA~\cite{Adloff:2001zj} that are not
included in the combined set; fixed-target Drell-Yan~ process \cite{Moreno:1990sf,Towell:2001nh,Webb:2003ps};
vector boson and inclusive jet production at the Tevatron~\cite{Abe:1996us,Acosta:2005ud,Abazov:2008qv,Abazov:2007pm,Abazov:2006gs,Aaltonen:2010zza,Aaltonen:2008eq,Abazov:2008ae}.
We also include inclusive jet production at the LHC~\cite{Aad:2011dm,Aad:2011fc},
which slightly reduces the uncertainty in the gluon PDF.

Depending on the candidate fit, the QCD coupling strength
was taken to be either the world average, $\alpha_{s}(M_{Z})=0.118$
\cite{Beringer:1900zz}, or a lower $\alpha_{s}(M_{Z})=0.115$, which
is preferred by the CT10 NNLO analysis when $\alpha_s(M_Z)$ is allowed to
vary. The factorization/renormalization
scale $\mu$ in DIS was set equal to the momentum transfer $Q$. To test
the sensitivity to the PDF parametrization form, the
initial scale $Q_{0}$ at which the input PDFs are provided was either
set to be $Q_{0}=1$ GeV independently of $m_{c}$ or
varied in the scan  together with $m_c$ as $Q_{0}=m_{c}-0.005$ GeV. 
Several forms of the gluon PDF parametrization were considered, since 
DIS charm production is sensitive to the gluon PDF $g(x,Q)$.
At the initial scale $Q_0$, we either constrained $g(x,Q_0)$ to be
positive at all $x$ or allowed it to be negative at small $x$,
provided the negative gluon did not lead to unphysical predictions. 
In the latter case of the negative gluon, the fit included
the H1 data on the longitudinal structure function 
$F_{L}(x,Q)$ \cite{Collaboration:2010ry}
and an additional theoretical constraint to ensure positivity of $F_{L}(x,Q)$
at $x>10^{-5}$. 

\subsection{Sensitivity of individual experiments}
\begin{figure}[tb]
\centering{}
\includegraphics[width=0.3\textwidth]{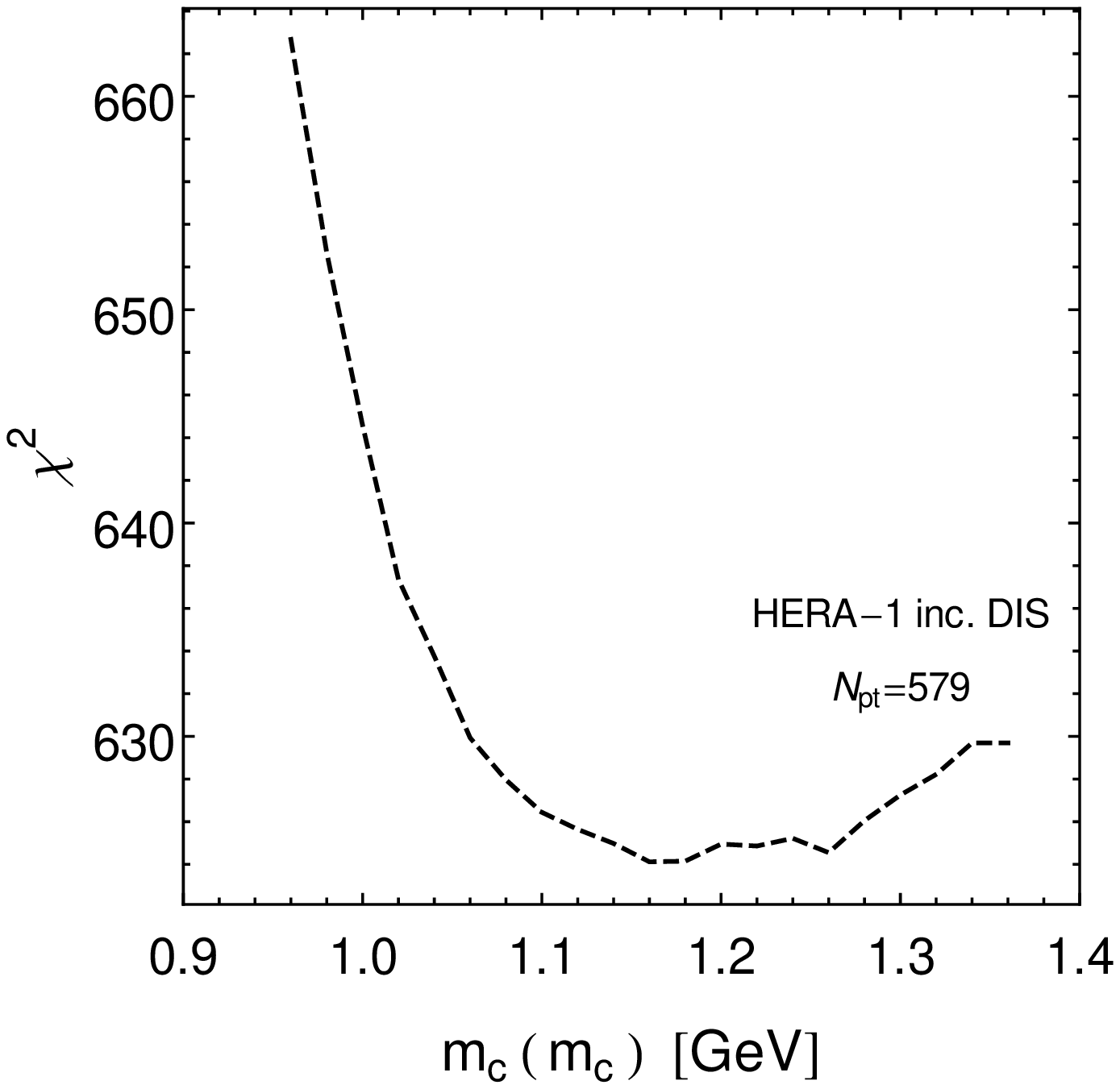}
\includegraphics[width=0.3\textwidth]{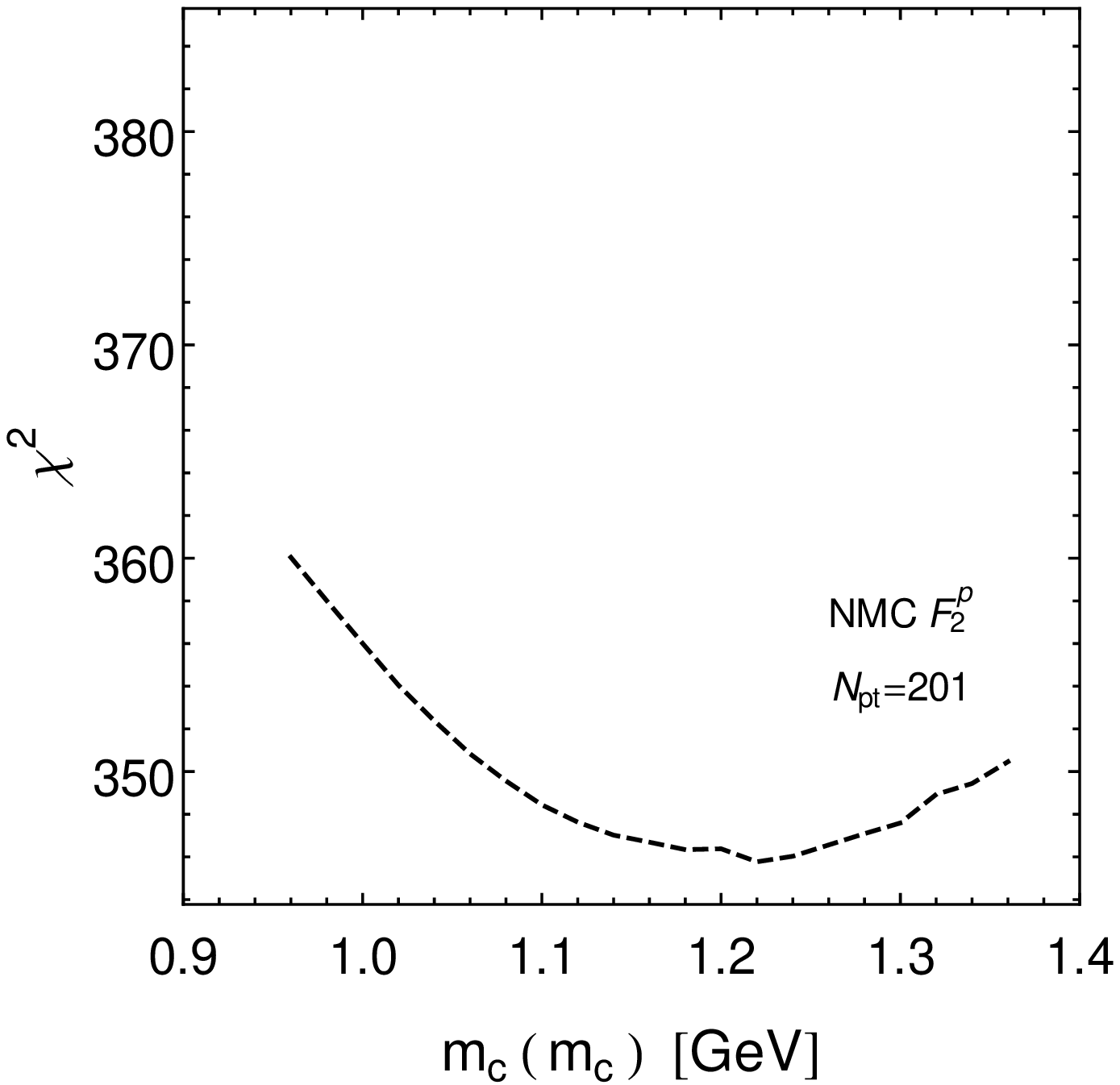}
\includegraphics[width=0.3\textwidth]{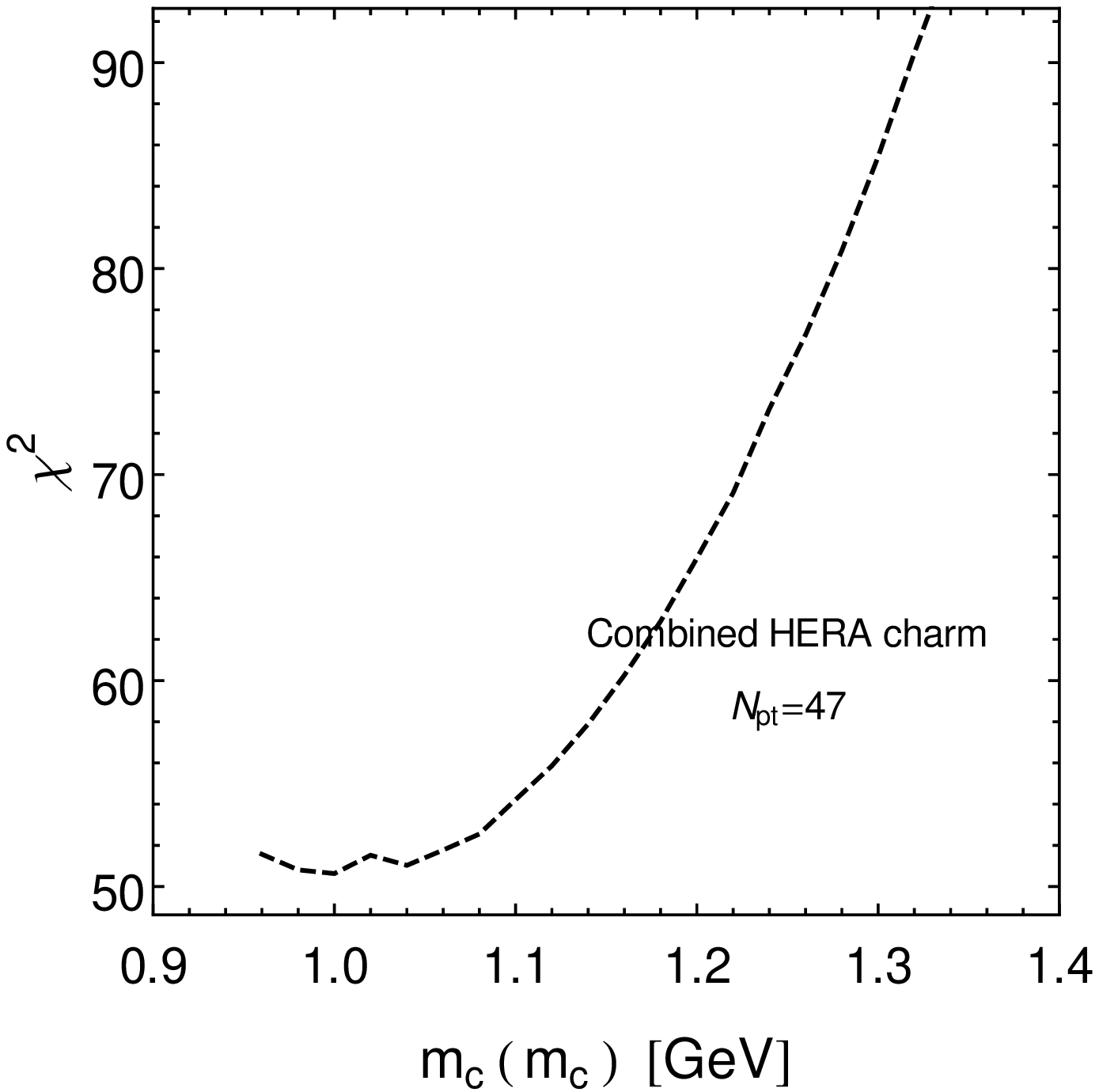}
\vspace{-1ex}
\caption{\label{fig:ind} Individual $\chi^{2}$ contributions of the combined
HERA-1 inclusive DIS data, NMC $F_{2}^{p}$ data, and combined HERA
charm quark production data as a function of $m_{c}(m_c)$.}
\end{figure}

Among all experiments included in the mass scan, the tightest constraints
on $m_{c}$ are imposed by the inclusive DIS HERA data \cite{Aaron:2009aa}
and HERA charm production \cite{Abramowicz:1900rp}. Weaker constraints
also arise from the other DIS experiments, 
notably the NMC measurement of $F_{2}^{p}(x,Q)$. 

Fig.~\ref{fig:ind} shows a sample of behavior of 
$\chi^{2}$  in each of three experiments 
as a function of $m_{c}$.
The inclusive DIS data from HERA (upper left inset) and NMC
(upper right inset) broadly prefer an $m_{c}$ in the range 1.1-1.3
GeV, while the charm HERA data (upper right inset) prefers lower
$m_{c}$ values of order 1 GeV. The NMC shown in the
lower inset prefers a higher $m_{c}$ of order 1.2 GeV in accord
with the inclusive HERA data, but with flatter $\chi^2$ dependence. 
While in the shown mass scan the inclusive DIS data
prefer a higher $m_{c}$ value than in DIS charm production, the 
$\chi^{2}$ minimum in inclusive DIS may shift to lower values of about
1.05-1.1 GeV depending on the parametrization forms of the PDFs and
other fit assumptions. 

We note that the semiinclusive reduced cross sections on HERA
charm production were derived from charm differential distributions
by applying a significant acceptance correction 
computed with the program \textsc{HVQDIS} \cite{Harris:1997zq} 
in the FFN scheme.  Due to the mismatch
in the schemes, the $m_{c}(m_{c})$ preferred by the HERA charm data
may be biased when determined in the S-ACOT-$\chi$ general-mass scheme.
Nevertheless, we see from Fig.~\ref{fig:ind} that 
the $m_{c}(m_{c})$ values that are separately preferred
by the inclusive DIS and charm DIS data sets are about the same.
In another cross check, we used separate, rather than combined,
data sets for HERA charm production 
\cite{Chekanov:2008yd,Chekanov:2009kj,Aaron:2009af,Aaron:2009jy,Aktas:2006py,Breitweg:1999ad,Chekanov:2003rb,Aaron:2011gp}.
Such fit did not differ much from the fit based on the combined
HERA charm data set.

\subsection{The global $\chi^2$ and PDF uncertainty \label{tol}}

\begin{figure}[tb]
\begin{centering}
\includegraphics[width=0.45\textwidth]{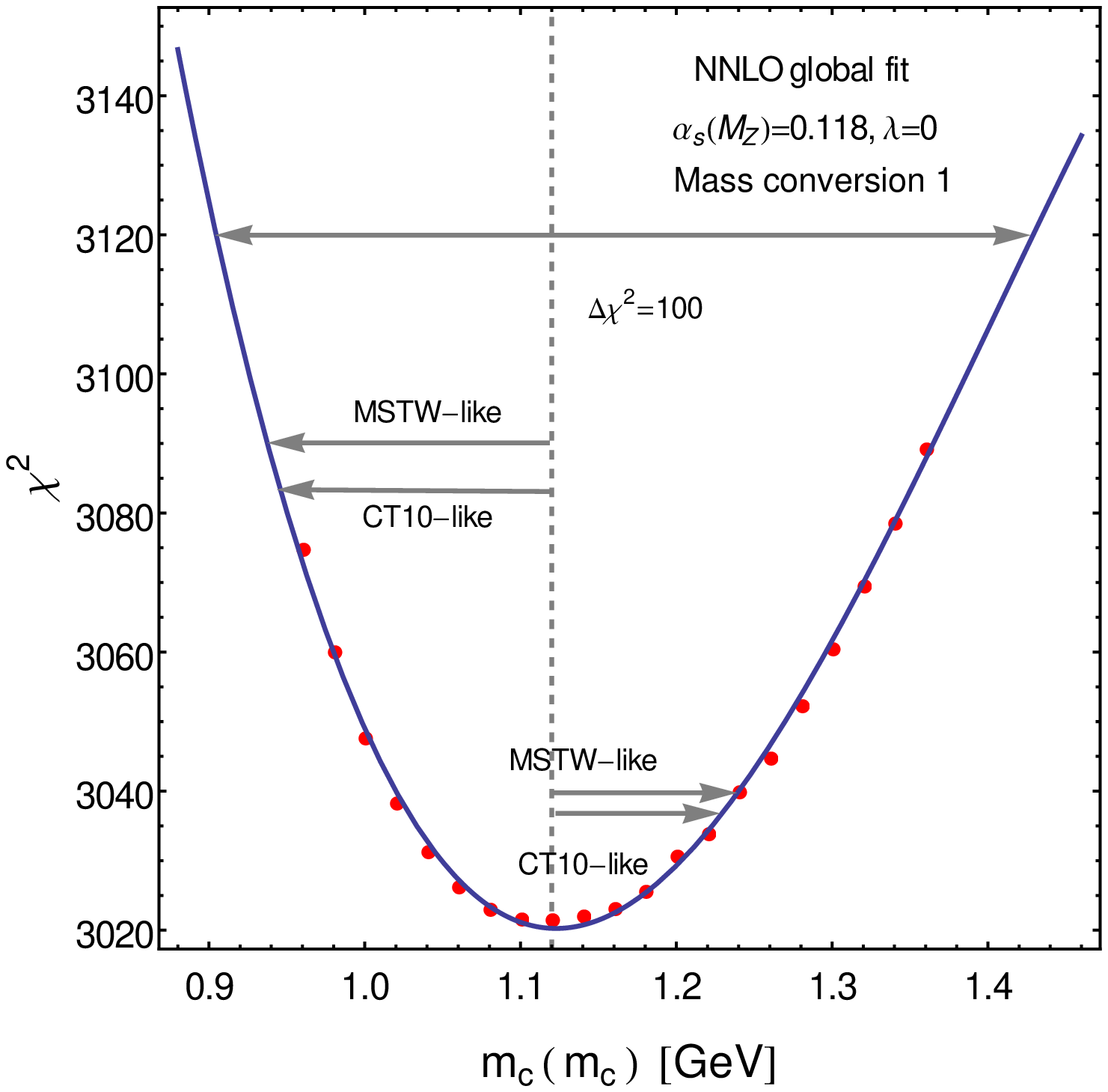}~\includegraphics[width=0.45\textwidth]{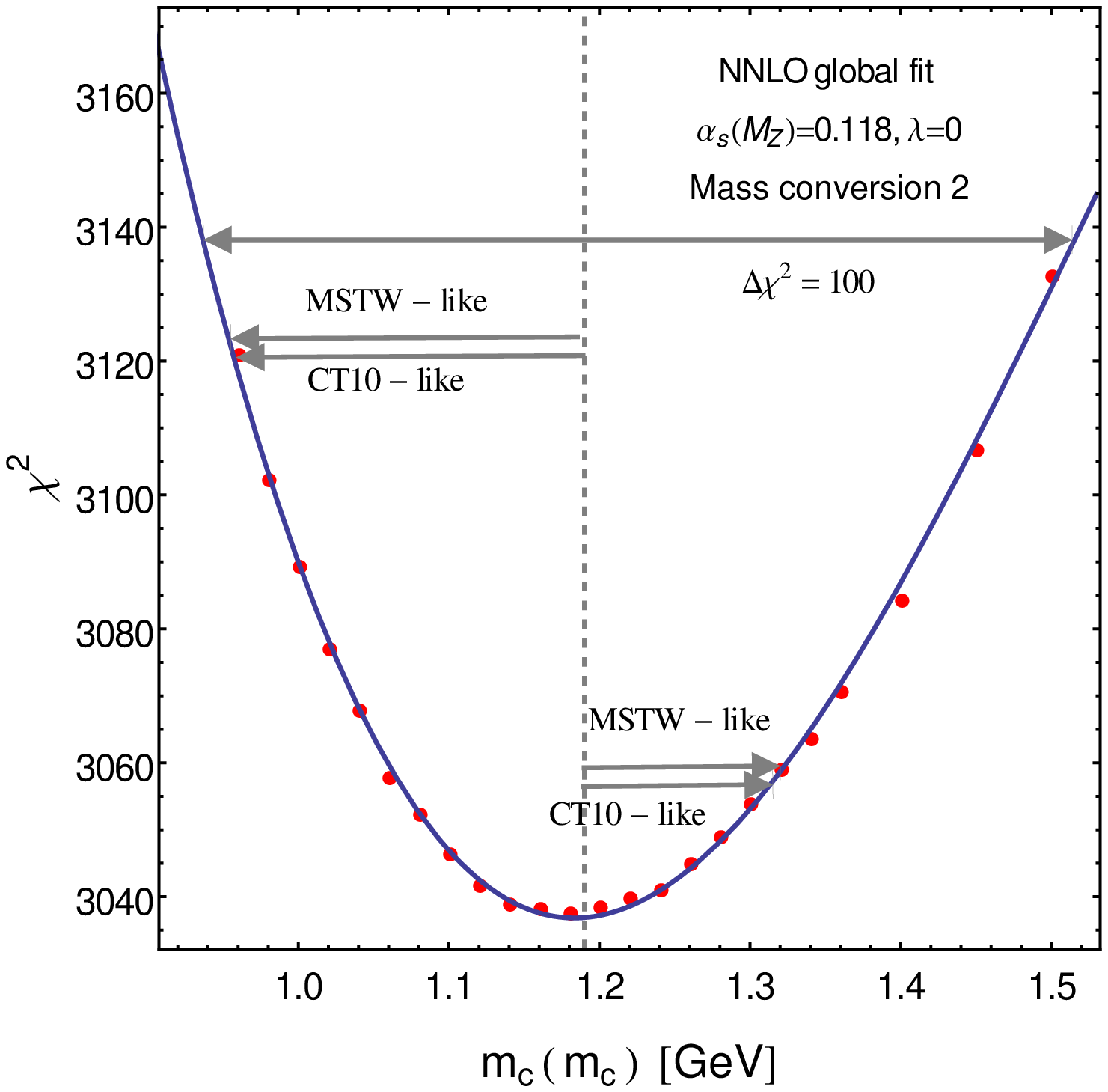} 
\par\end{centering}

\vspace{-1ex}
 \caption{\label{gchi2} Global $\chi^{2}$ of the S-ACOT-$\chi$ NNLO fit as
a function of the $\overline{{\rm MS}}$ charm mass. Lines with left/right
arrows indicate 90\% C.L. intervals obtained with different tolerance
criteria.}
\end{figure}

The constraints from
various experiments are generally compatible and 
produce a well-defined minimum in the global $\chi^2$.
The plots of $\chi^{2}$ for all experiments 
in the above $m_{c}(m_{c})$ scans is shown in Fig.~\ref{gchi2}
for the full $\overline{\rm MS}\rightarrow\mbox{pole}$ mass 
conversion (left inset) and truncated conversion (right inset). 
The PDF parameters were refitted for every
$m_{c}(m_c)$. The functional form for $\chi^2$ can be found
in Ref.~\cite{Gao:2013xoa}. The default 
rescaling parameter $\lambda=0$ was used. The scattered points 
are for individual fits at discrete $m_c(m_c)$ values, while the
continuous line indicates smooth interpolation across the individual
fits.  

In the figure, the preferred value of the $\overline{{\rm MS}}$ charm
mass, corresponding to the minimum of $\chi^{2}$, is $m_{c}(m_{c})=1.12$
GeV for the full conversion and 1.18 GeV for the truncated conversion. 
The optimal charm mass is below the world-average 
$m_{c}(m_{c})=1.275\pm0.025$ GeV, 
and this general trend is observed in all fits. 

From the mass dependence of $\chi^{2}$ in the figure, we can compute
the uncertainty on $m_c(m_c)$ due to the PDF parameters. 
Since the CT analysis traditionally operates with the 90\% confidence
level (90\% C.L.) uncertainty, 
we compared 3 different criteria
for defining it for  $m_{c}(m_{c})$: 1) the uniform $\chi^{2}$ tolerance,
in which one assigns a 90\% C.L. to a $\Delta\chi^{2}\leq100$ variation
as in CTEQ6  \cite{Pumplin:2002vw}; 2) the
CT10-like criterion, which supplements the uniform $\chi^{2}$ tolerance
condition by additional $\chi^{2}$ penalties to prevent strong disagreements
with individual experiments \emph{on average} \cite{Lai:2010vv};
3) and the MSTW-like criterion, which does not introduce the uniform tolerance,
but requires the $\chi^{2}$ value for \emph{every} individual experiment
to lie within the specified confidence interval. In the latter two
methods, deviations of the PDF parameters 
are additionally constrained so as not to trigger a strong disagreement 
with one of the fitted experiments. This condition effectively 
reduces the PDF uncertainty in methods 2 and 3 
compared to the uniform tolerance criterion (method 1), 
see further details in \cite{Lai:2010vv}.

The PDF uncertainties on $m_{c}(m_{c})$ obtained with the three definitions
are reported in Table~\ref{TAB1}. The uncertainty according to the
uniform tolerance definition is larger than with the other two definitions,
as anticipated. 

\begin{table}[h!]
\begin{centering}
\begin{tabular}{|c|c|}
\hline 
\multicolumn{2}{|c|}{PDF uncertainty $\delta m_{c}$ {[}GeV{]} (90\% C.L.)}\tabularnewline
\hline 
$\Delta\chi^{2}\leq100$  & $\delta m_{c}=_{-0.22}^{+0.30}$ \tabularnewline
\hline 
CT10-like  & $\delta m_{c}=_{-0.17}^{+0.11}$\tabularnewline
\hline 
MSTW-like  & $\delta m_{c}=_{-0.18}^{+0.12}$\tabularnewline
\hline 
\end{tabular}
\par\end{centering}

\caption{The PDF uncertainty on the optimal value of $m_{c}$ extracted from
the fit at the 90\% C.L. by using three different criteria.}

\label{TAB1} 
\end{table}

\begin{figure}[tb]
\begin{centering}
\includegraphics[width=0.5\textwidth]{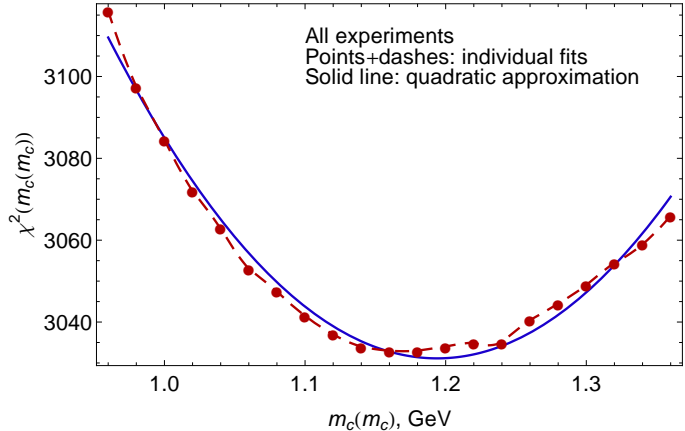} 
\par\end{centering}
\vspace{-1ex}
\caption{\label{fig:dchi}Total $\chi^2$ and its parabolic fit as functions
the charm mass $m_c(m_c)$.}
\end{figure}

The rest of the paper interchangeably operates with the 68\% and 
90\% C.L. intervals, with the former taken to be 
1.65 times smaller than the latter. 
It is insightful to 
compare the above tolerance criteria to the
procedure in the FFN analyses,
\cite{Alekhin:2012vu,Abramowicz:1900rp} that assigns the 68\%
C.L. PDF uncertainty to the increase in
the global $\chi^2$ by one unit. This 
definition of the PDF error \cite{Collins:2001es,BevingtonRobinson}
is applicable under ideal conditions and
leads to a smaller PDF error compared to the CT10 criterion.
We have examined the $\Delta\chi^{2}=1$ criterion as an alternative to
the CT10 criterion and 
found that it does not realistically describe the observed 
probability distribution. The reason is that the $\Delta\chi^2 =1$ criterion is
strictly valid when the experimental errors are Gaussian, which
implies quadratic dependence of  $\chi^2$ on $m_c(m_c)$.
The actual $\chi^2$ distribution is not perfectly
Gaussian and exhibits asymmetry as well as some random fluctuations. 

In Fig.~\ref{fig:dchi} we compare the observed distribution
of $\chi^2(m_{c,i})$ (scattered circles) 
with a fit by a second-degree polynomial 
$\chi^2_{parabola}(m_c)=A m_c^2+B m_c+ C$ shown by the solid
line. Deviations from the perfect quadratic behavior can be
characterized by
\begin{equation}
X^2\equiv\sum_{i=1}^{N_{m_c}}(\chi^2_{parabola}(m_{c,i})-\chi^2(m_{c,i}))^2,
\label{X2}
\end{equation}
where $N_{m_c}$ is the number of discrete $m_c(m_c)$ values in the
scan. In the ideal Gaussian case, when the individual $\chi^2(m_{c,i})$
follow neatly the parabola in the scanned $m_c(m_c)$ region, 
$X^2$ is much less than  $N_{m_c}-1$, and $\delta m_c=\sqrt{1/A}$ (about 0.025
GeV in our fits) provides a fair estimate of the $1\sigma$ error. 
But in the actual fits of the kind indicated by the circles in
Fig.~\ref{fig:dchi}, $X^2/(N_{m_c}-1)$ is of order 2.5, hence $\sqrt{1/A}$
underestimates the $m_c$ error by a factor of about 2.5. This 
reflects the probability distribution that is broader 
than the ideal parabola and hence contains less than 68\% of the net
probability in the $m_c$ region where the parabolic growth results in
$\Delta\chi^2=1$. 

A partial remedy is procured by symmetric rescaling, if one
defines $\widetilde X^{2}\equiv X^2/C$  with a constant $C\approx 2.5$ so that
$\widetilde X^{2}\approx 1$
for the observed distribution of $\chi^2(m_{c,i})$. When the PDF
uncertainty is derived from the rescaled $\widetilde X^{2}$ statistic, 
it is larger by a factor $C$ compared to the idealized assumption. 
Symmetric rescaling increases the 68\% C.L. error but does not fix the
shape of $\chi^2(m_c)$. After the rescaling, the PDF error of 0.06 GeV 
gets closer to the one obtained by the CT10-like criterion, 
which is of order 0.09 GeV at 68\% C.L. if the asymmetric errors 
are averaged over.\footnote{A broadscale argument is also available that the probability distribution ${\cal P}(m_c)\propto \exp\left(-\chi^2(m_c)/2\right)$ on which the $\Delta \chi^2=1$ criterion is based underestimates the confidence levels in PDF fits \cite{Ball:2011gg}.} 

Our main conclusion is that, in the contest of our NNLO 
extraction of $m_c(m_c)$, the quadratic assumption fails to
describe the actual $\chi^2(m_c(m_c)$ distribution to the extent needed to
justify the $\Delta\chi^2=1$ prescription. The actual distribution is
flatter near the minimum, asymmetric, and has occasional
fluctuations. A number of reasons can explain this
behavior. It has been observed in the earlier work
\cite{Lai:2010vv,Pumplin:2009nm,Pumplin:2009sc} that moderate
disagreements between the fitted experiments broaden 
the probability distribution around the global minimum of $\chi^2$ compared to the ideal Gaussian case. The net effect of these disagreements may be approximated, to the first order, by increasing the $\Delta \chi^2=1$ PDF error by a factor
of two or three, {\it i.e.}, about the same factor as in the $m_c(m_c)$ fit.

\subsection{Systematic uncertainties}
\begin{table}[h!]
\begin{centering}
\begin{tabular}{c|c|c|c|c|c}
\hline 
Theoretical systematic uncertainty   & $m_c^{pole}$ conversion & DIS scale  & $\alpha_{s}(M_{Z})$  & $\lambda$  & $\chi^{2}$ definition \tabularnewline
\hline 
Parameter range  & -- & $[Q/2,\ 2Q]$  & {[}0.116,\, 0.120{]}  & {[}0, 0.2{]}  & -- \tabularnewline
\hline 
$\delta m_c(m_c)$ (GeV) & 0.07  & ${}^{+0.02}_{-0.02}$  &
${}^{+0.01}_{-0.01}$  & ${}^{+0.14}_{-0}$  & 0.06  \tabularnewline
\hline 
\end{tabular}
\par\end{centering}

 \caption{\label{unc} Shifts of the optimal value of the charm mass $m_{c}(m_{c})$
obtained by varying theoretical inputs.}
\end{table}

In addition to the PDF uncertainty associated with
experimental errors, 
Table \ref{unc} summarizes systematic uncertainties on $m_{c}(m_c)$ associated
with theoretical inputs, including the
$\overline{\rm MS}\rightarrow\mbox{pole}$ conversion procedure, 
the factorization/renormalization
scale, $\alpha_{s}(M_{Z})$, the $\lambda$ parameter in the rescaling
variable, and implementation of experimental correlated systematic
errors. The last source of uncertainty arises from the
existence of several prescriptions for including correlated systematic
errors from the fitted experiments into the figure-of-merit function
$\chi^{2}$ \cite{Ball:2012wy,Gao:2013xoa}. Depending on the prescription,
the relative correlated errors published 
by the experiments can be interpreted as fractions 
of either central data values or theoretical
values. These methods are
designated as ``extended T'' and ``D'' methods in Ref.~\cite{Gao:2013xoa}.
This leads to numerical differences in absolute correlated
errors, which may affect the outcomes of the fit, as 
discussed in the above references. We estimate the associated
uncertainty by alternating between the two normalization methods 
for the correlation matrices of the DIS processes. 

\begin{figure}[h!]
\begin{centering}
\includegraphics[width=0.5\textwidth]{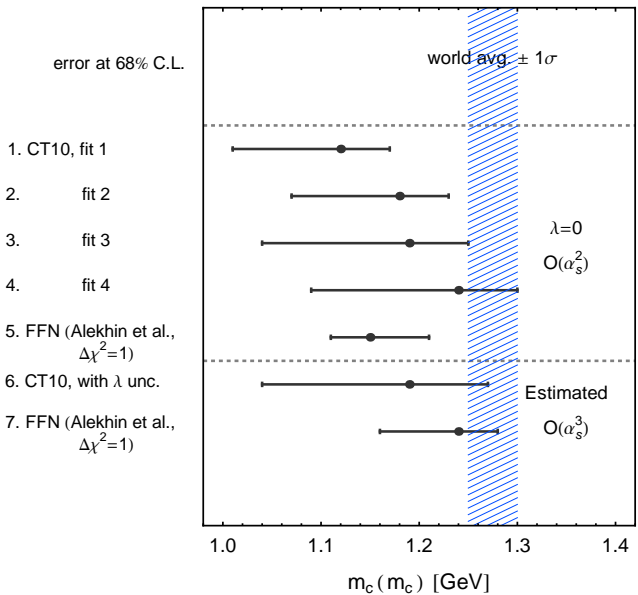} 
\par\end{centering}

\vspace{-1ex}
\caption{\label{fig:bench} Comparison of the $\overline{\textrm{MS}}$ charm mass 
and its uncertainty extracted with various methods. 
Illustrated are also $m_c(m_c)$ values obtained in an FFN analysis 
by Alekhin et al.~\cite{Alekhin:2010sv}.}
\end{figure}

The values of $m_c(m_c)$ obtained under various assumptions are
illustrated by Fig.~\ref{fig:bench}. At order $\alpha_s^2$, 
the highest fully implemented order in
our calculation, we show $m_c(m_c)$ found with four methods.
Methods 1 and 2 correspond to
the ``extended $T$'' and ``experimental'' $\chi^2$ definitions
respectively~\cite{Gao:2013xoa},   
both using the full $\overline{{\rm MS}}\rightarrow\mbox{pole}$ mass conversion formula, 
and $\lambda=0$. Methods 3 and 4 are the same as 1 and 2, but with truncated  
mass conversion equivalent to computing the coefficient functions in the
$\overline{\rm MS}$ scheme.
The resulting $m_c(m_c)$ values in the four methods are 
$1.12^{+0.05}_{-0.11}$, $1.18^{+0.05}_{-0.11}$, 
$1.19^{+0.06}_{-0.15}$ and $1.24^{+0.06}_{-0.15}$ GeV,
respectively. Here and in the figure 
we quote the  68\% C.L. PDF uncertainties defined
as in the CT10 analysis, cf. Sec.~\ref{tol}.

\begin{figure}[h!]
\begin{centering}
\includegraphics[width=0.4\textwidth]{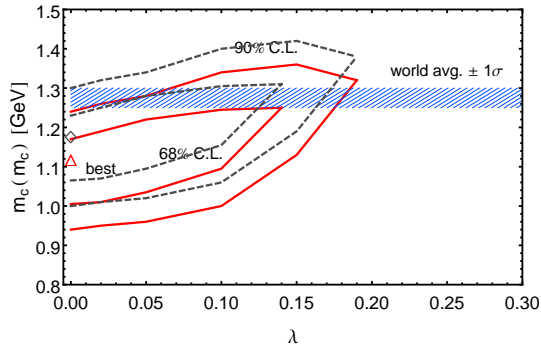} 
\par\end{centering}

\vspace{-1ex}
\caption{\label{fig:con} Preferred regions for $m_{c}(m_{c})$ vs. the rescaling
parameter $\lambda$. The best-fit values and confidence intervals
are shown for two alternative methods
for implementation of correlated systematic errors.}
\end{figure}

The dependence on the $\lambda$ parameter is illustrated by 
Fig.~\ref{fig:con}, showing boundaries of the 68\% and 90\%
C.L. regions when $\lambda$ takes values 
on the horizontal axis, for the 
``extended $T$'' (solid lines) and ``experimental''
(dashed lines) definitions of $\chi^2$, and using the full 
$\overline{\rm MS}\rightarrow\mbox{ pole}$ conversion.\footnote{Similar $\lambda$ dependence is observed for the truncated conversion.} 
The red empty triangle and black diamond symbols 
are the best-fit values of $m_{c}(m_{c})$ obtained with the two
$\chi^2$ prescriptions, equal to  1.12 and 1.18 GeV, and reached
when $\lambda\approx 0$ in both cases.
Values of $\lambda$ above 0.14 and 0.20 are disfavored at 
68\% (90\%) C.L.\footnote{The 
CT10 or MSTW-like tolerance criteria lead to about the same boundaries.}
Finally, the horizontal blue band indicates the world-average
interval $m_{c}(m_{c})=1.275\pm 0.025$ GeV. 

As we see, there is some spread in the $m_c$ values depending on the
adopted $\overline{\rm MS}\rightarrow\mbox{pole}$ conversion and $\chi^2$
definition. In addition, moderate dependence exists on the rescaling
parameter $\lambda$, associated with missing higher-order corrections.
We can estimate the projected range for the ${\cal O}(\alpha_s^3)$
value of $m_c(m_c)$ in method 3 (or any other method) 
according to the CT10-like criterion
from the $\chi^2$ dependence for a range of $\lambda$ values. 
This produces 
$1.19^{+0.08}_{-0.15}$ GeV for the estimated 
${\cal O}(\alpha_s^3)$ value in method 3, 
as shown in line 6 in Fig.~\ref{fig:bench}. Here the errors are
estimated from the 68\% C.L. contour for $\chi^2$ vs. $\lambda$, and
the scale and $\alpha_s$ errors are added in
quadrature.

The central $m_c(m_c)$ is consistent with the PDG value of $1.275\pm 0.025$
GeV \cite{Beringer:1900zz} 
within the errors. A tendency of the fits to undershoot the PDG
value may be attributable to the missing ${\cal O}(\alpha_s^3)$
contribution \cite{Alekhin:2012vu}. The results of our fit are
compatible with $m_c(m_c)$  determined 
in the (FFN) scheme \cite{Alekhin:2012vu} both at the exact ${\cal O}(\alpha_s^2)$ and approximate ${\cal O}(\alpha_s^3)$, cf. lines 5 and 7 in Fig.~\ref{fig:bench}. However, our PDF error of about 0.08 GeV 
is 2.7 times larger than the one quoted in the FFN study. The
main reason is that the 68\% C.L. PDF 
uncertainty of the FFN analysis corresponds to $\Delta\chi^{2}=1$ 
and hence is smaller than the CT10-like uncertainty. As discussed above, 
our $\chi^2$ dependence on $m_c(m_c)$ is not compatible with the
ideal Gaussian behavior and could be accommodated by increasing the
PDF error by a factor 2-3 compared to the $\Delta\chi^{2}=1$
definition. Besides this difference in the PDF
uncertainty, the results for $m_c(m_c)$ from the S-ACOT-$\chi$ and FFN
fits are in general agreement.

\begin{figure}[h]
\begin{centering}
\includegraphics[width=0.48\textwidth]{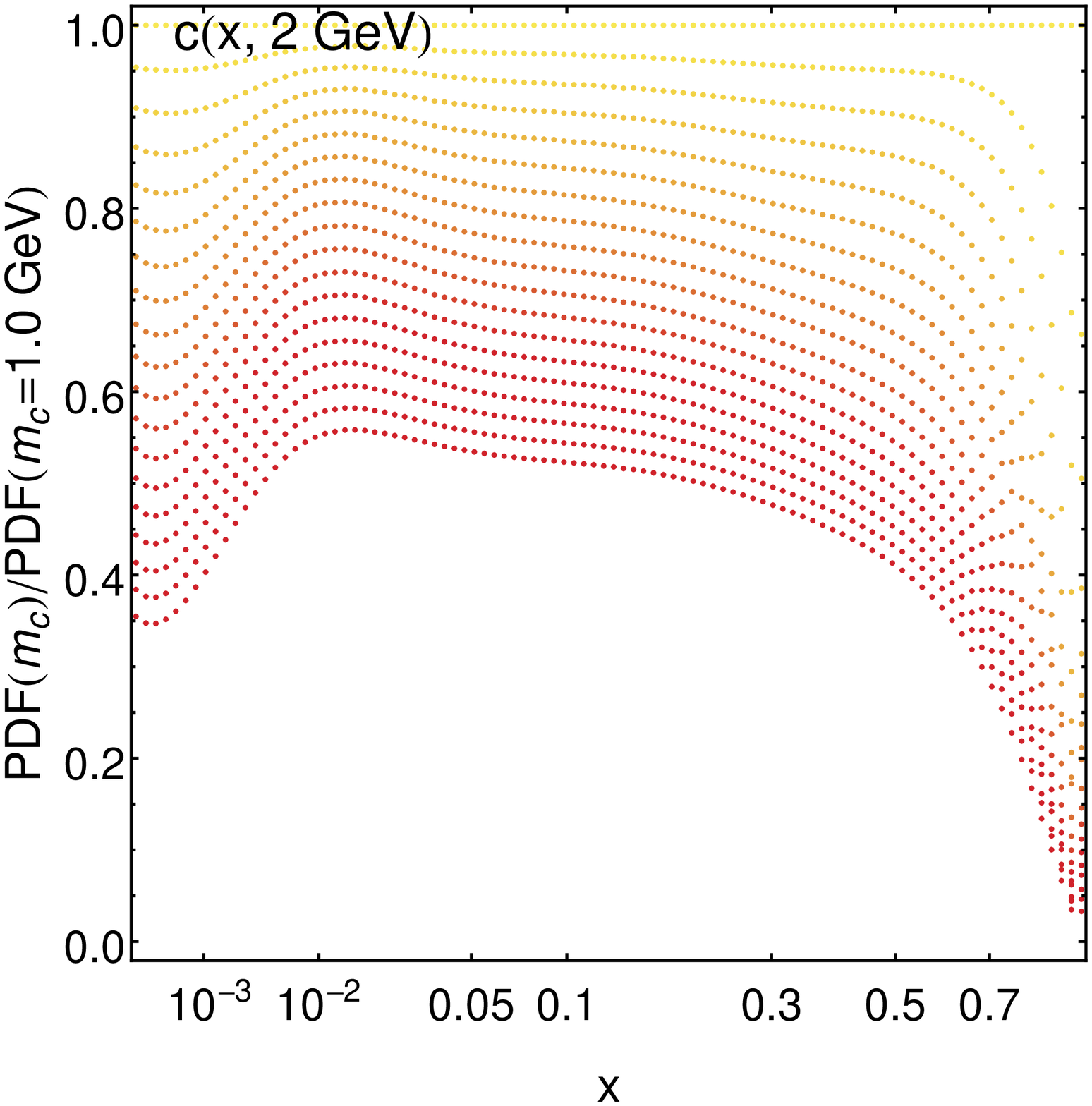}\includegraphics[width=0.48\textwidth]{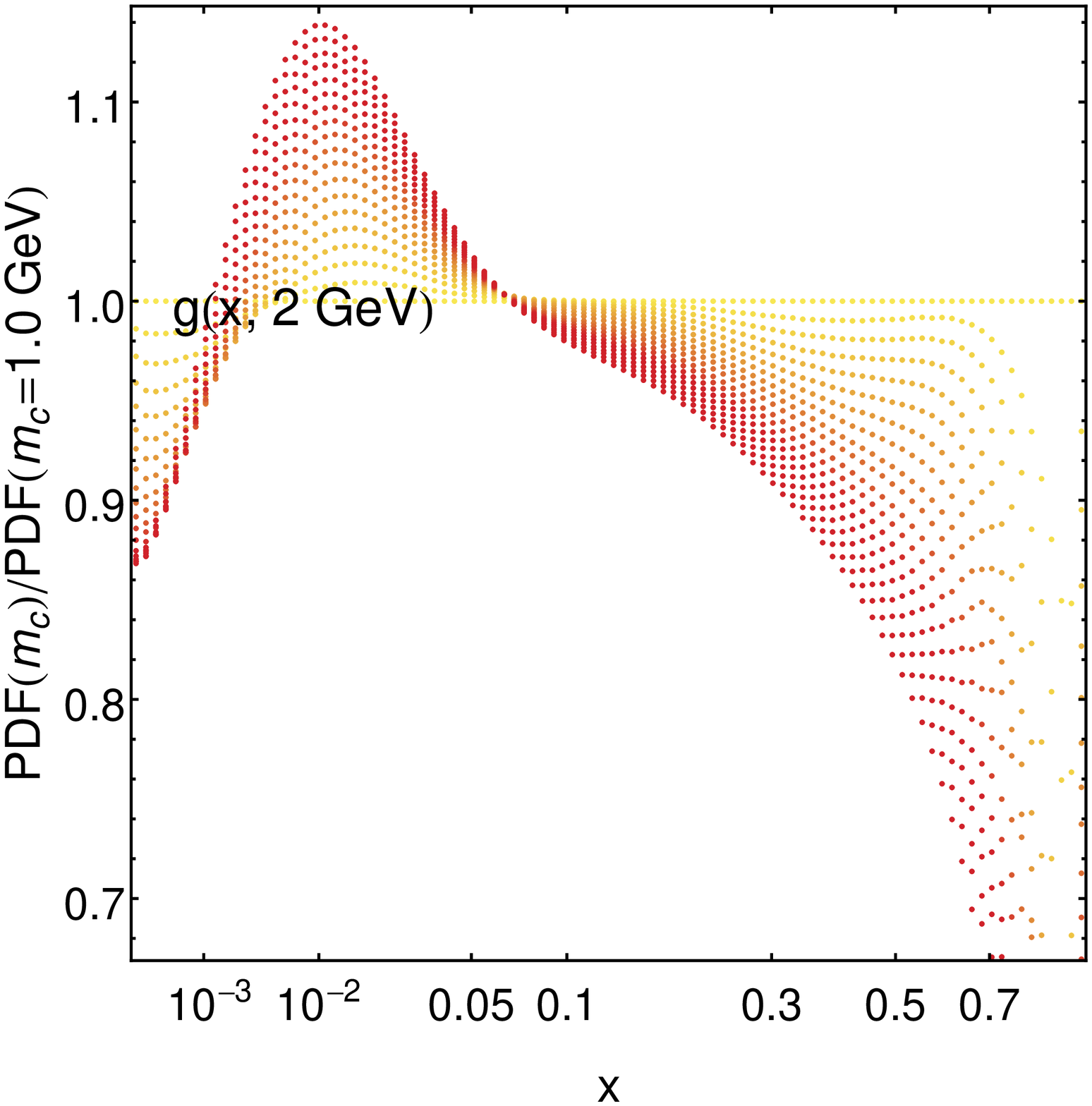}\\
\includegraphics[width=0.48\textwidth]{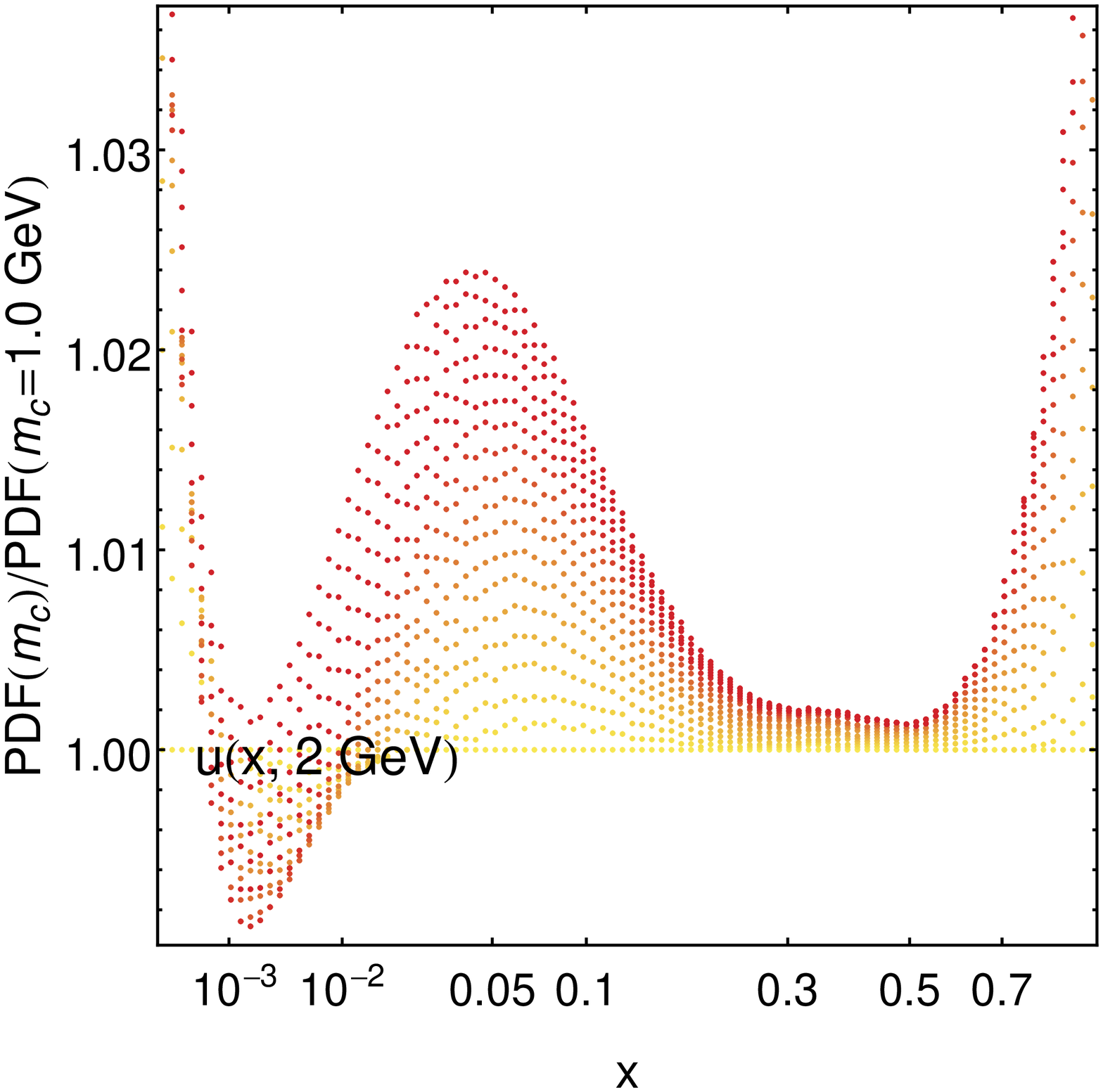}\includegraphics[width=0.48\textwidth]{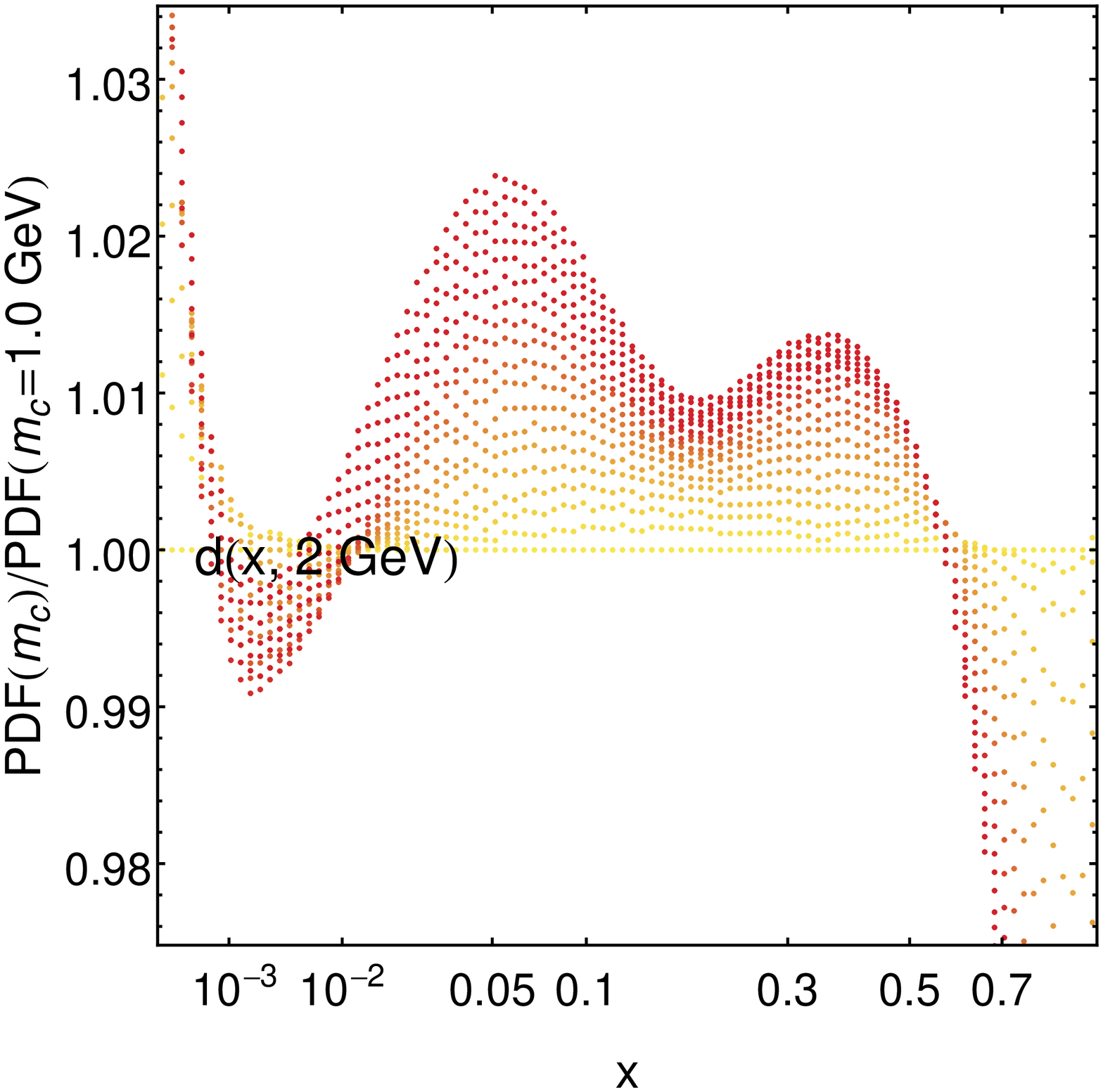} 
\par\end{centering}

\vspace{-1ex}
 \caption{\label{fig:pdf} Relative changes in select PDFs $f_{a/p}(x,Q)$ at
$Q=2$ GeV obtained in a series of PDF fits with $m_{c}(m_{c})$ ranging
from 1 to 1.36 GeV, plotted as ratios to the respective PDFs for $m_{c}(m_{c})=1$
GeV. Default rescaling ($\lambda=0$) in DIS coefficient functions
is assumed. Darker colors correspond to larger $m_{c}(m_{c})$ values.}
\end{figure}

\subsection{Implications for PDFs and collider observables\label{sec:Implications}}

As $m_{c}(m_{c})$ changes in the mass scan, parametrizations of
the PDFs are adjusted so as to maximize agreement with the data. Representative
best-fit PDFs from a charm mass scan are plotted in
Fig.~\ref{fig:pdf} at $Q=2$ GeV. In this example, we vary the charm mass in the
interval $1.0\leq m_{c}(m_{c})\leq1.36$ GeV that is about the same
as the 90\% C.L. CT10-like PDF uncertainty determined in the
previous section. For each $m_{c}(m_{c})$ from this range,
we refit the PDFs, while keeping the other theoretical inputs at their
default values, in particular assuming $\alpha_{s}(M_{Z})=0.118$
and $\lambda=0$ in the rescaling variable. Darker color points in
Fig.~\ref{fig:pdf} correspond to larger mass values. 

In general, as the charm mass is increased, both the charm PDF (upper
left subfigure) and charm contributions to DIS cross sections are
suppressed. Consequently, the gluon PDF (upper right subfigure)
is enhanced in the intermediate $x$ region, $10^{-3}\sim10^{-2}$
so as to partly compensate for this reduction. This is accompanied
by moderate enhancements in the up and down quark PDFs (lower subfigures)
at $x\approx10^{-2}-0.5$ and slight suppression of the same PDFs at
$x\approx10^{-3}$.

\begin{figure}[tbp]
\begin{centering}
\includegraphics[width=0.4\textwidth]{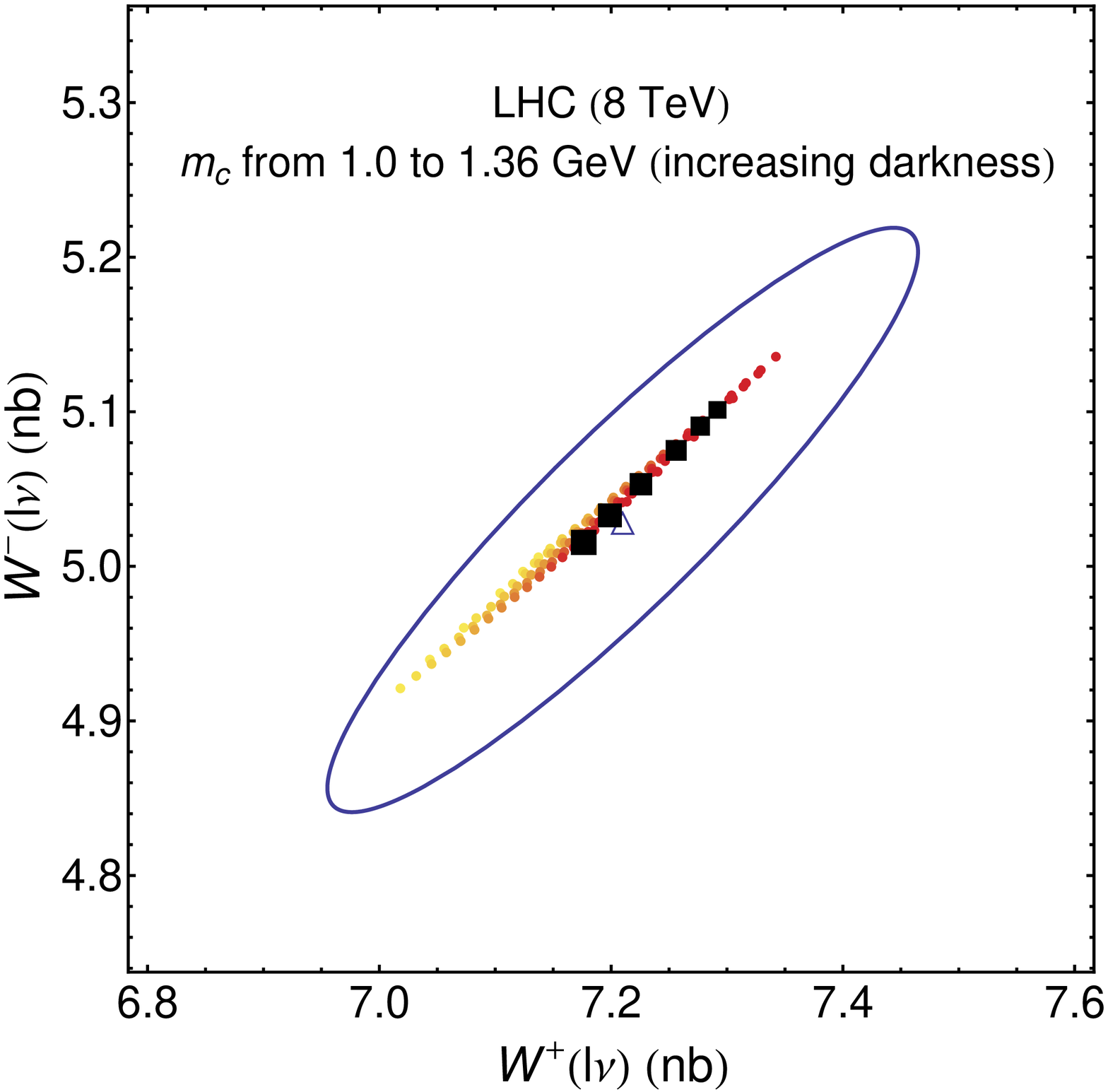}
\includegraphics[width=0.41\textwidth]{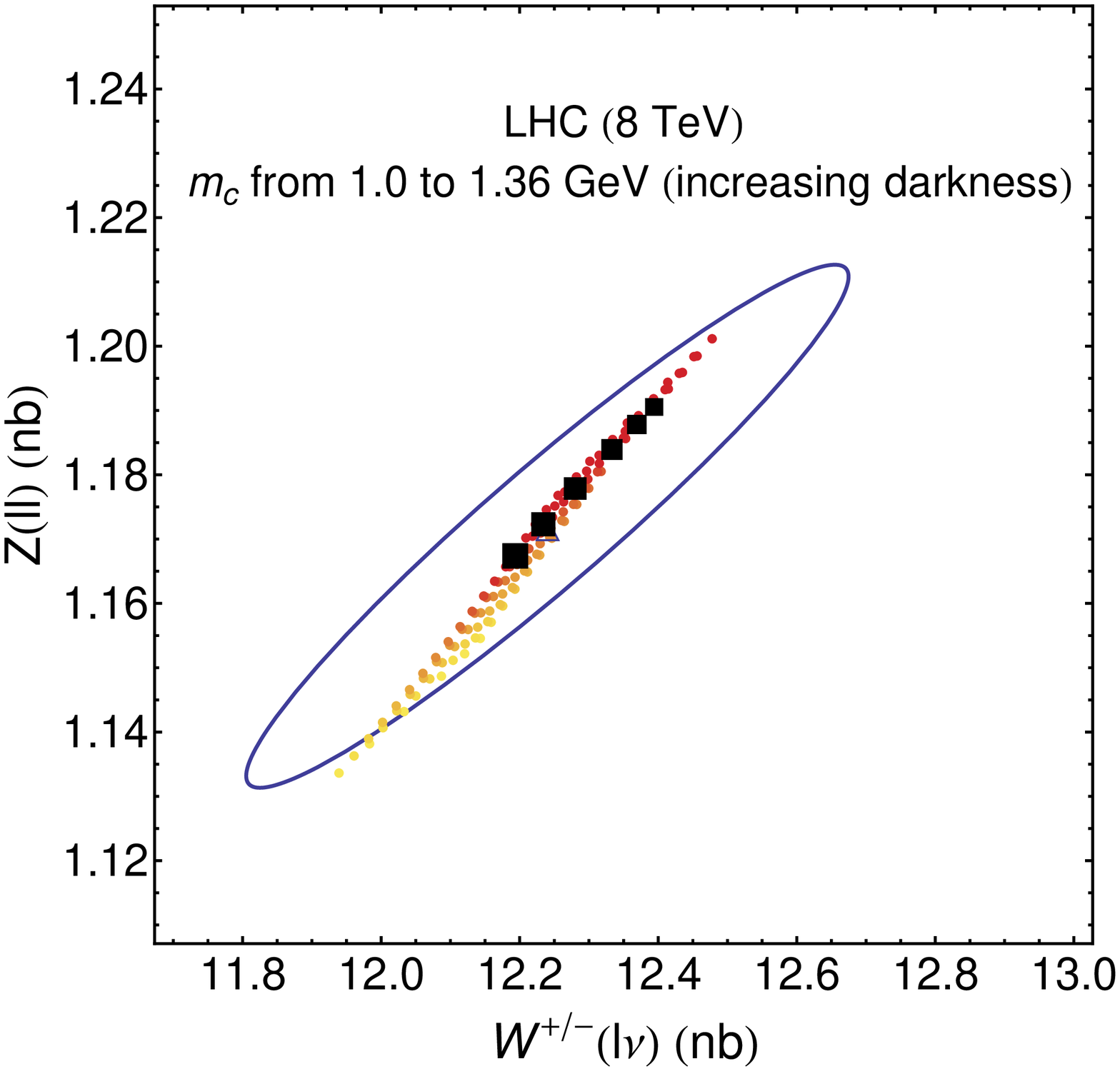}\\
\vspace{0.3in}
\includegraphics[width=0.4\textwidth]{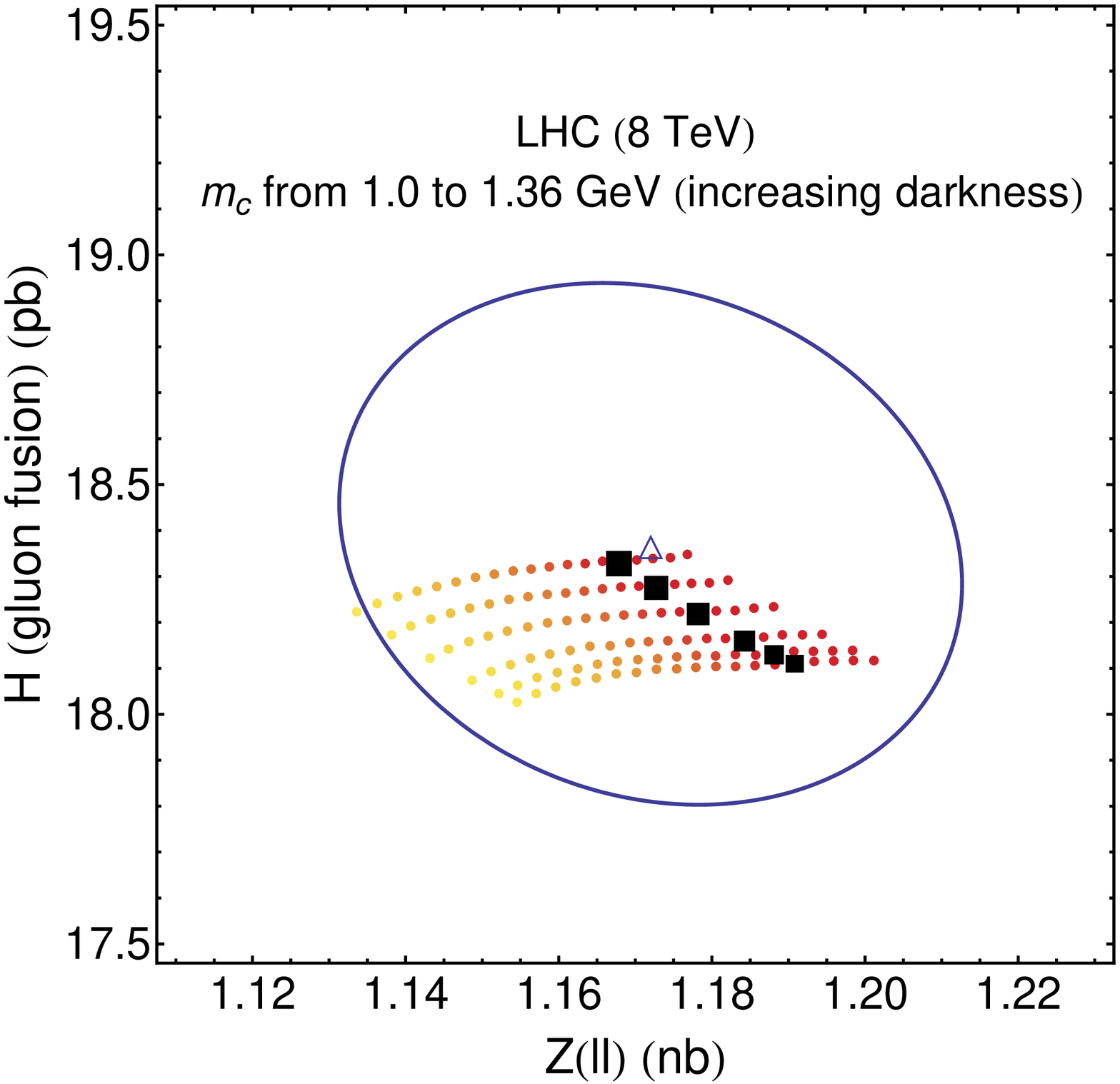} 
\includegraphics[width=0.4\textwidth]{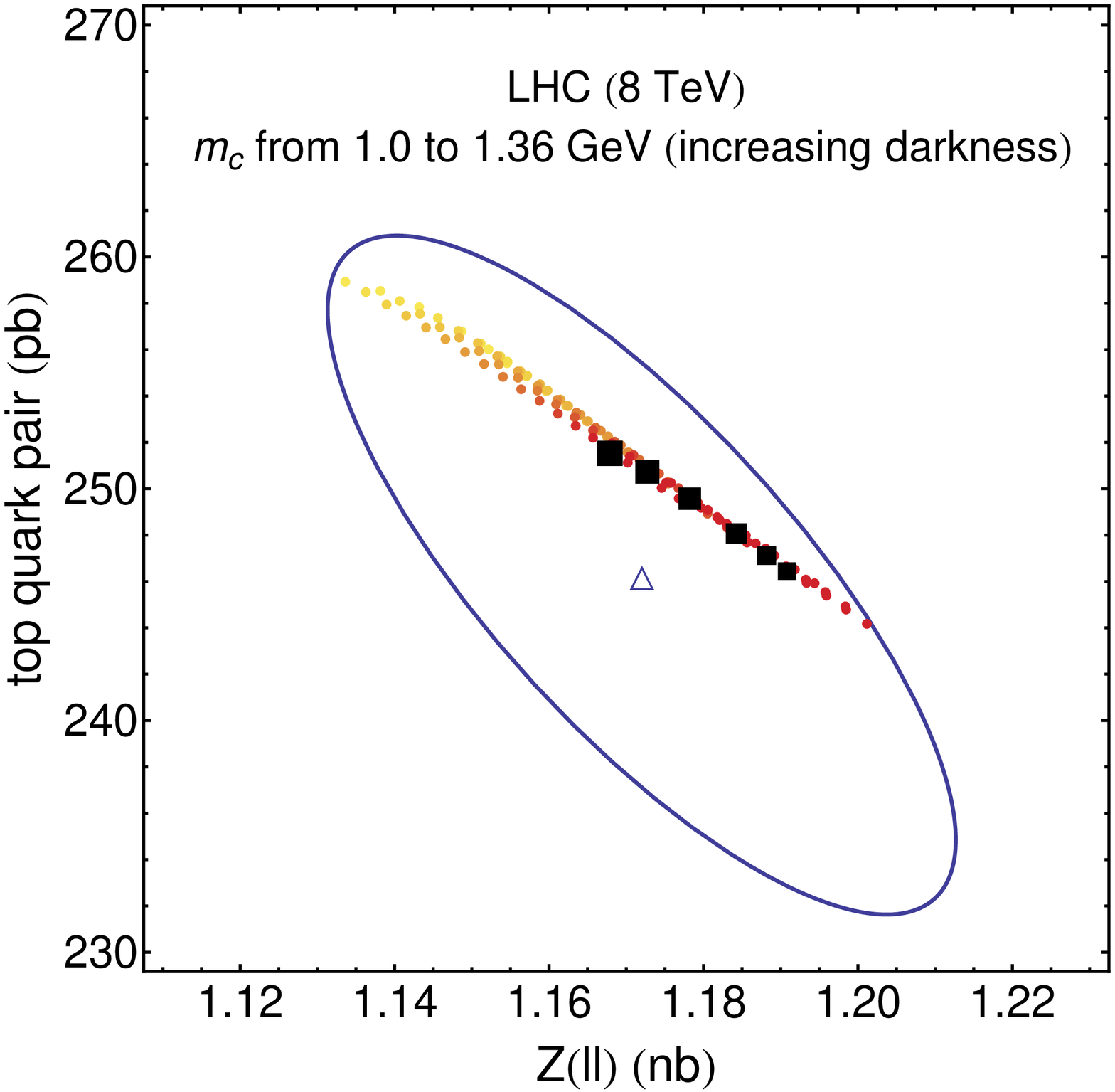} 
\par\end{centering}
\vspace{-1ex}
 \caption{\label{fig:wz1} Plot of NNLO cross sections for $W^{\pm}$, $Z^{0}$,
Higgs boson production through gluon fusion, and top quark pair production
at the LHC (8 TeV) for charm quark mass
$m_{c}(m_{c})$ ranging from 1 to 1.36 GeV and $\lambda=\{0,\ 0.02,\ 0.05,\ 0.1,\ 0.15,\ 0.2\}$.
Darker color corresponds to larger mass values. The black boxes
represent the cross sections evaluated by using $m_{c}(m_{c})=1.28$ (close to world average)
GeV and the explored $\lambda$ values. The empty triangle and ellipse
indicate the central predictions and its 90\% C.L. interval based on the CT10NNLO fit.}
\end{figure}

\begin{figure}[tbp]
\begin{centering}
\includegraphics[width=0.4\textwidth]{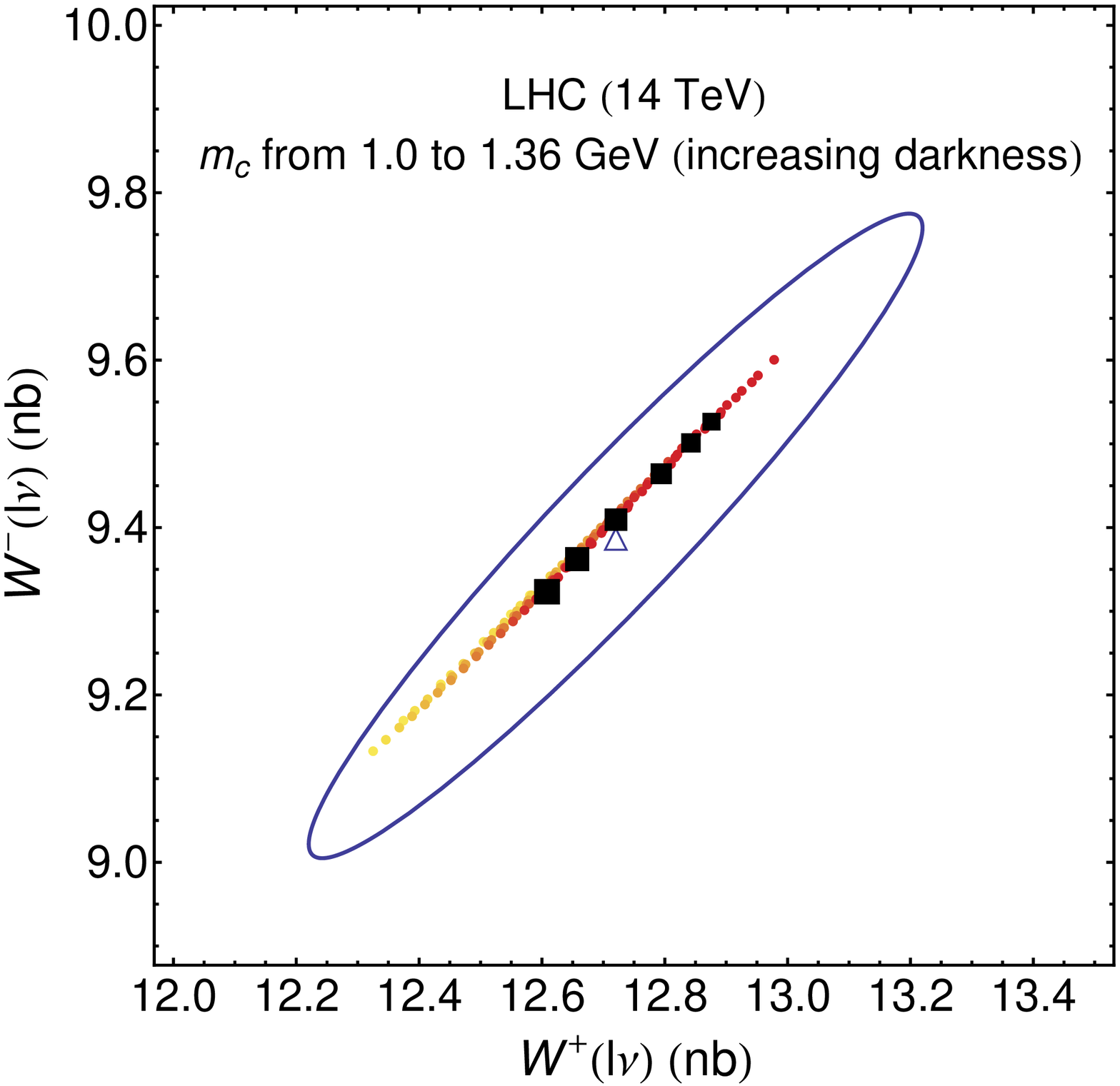}
\includegraphics[width=0.4\textwidth]{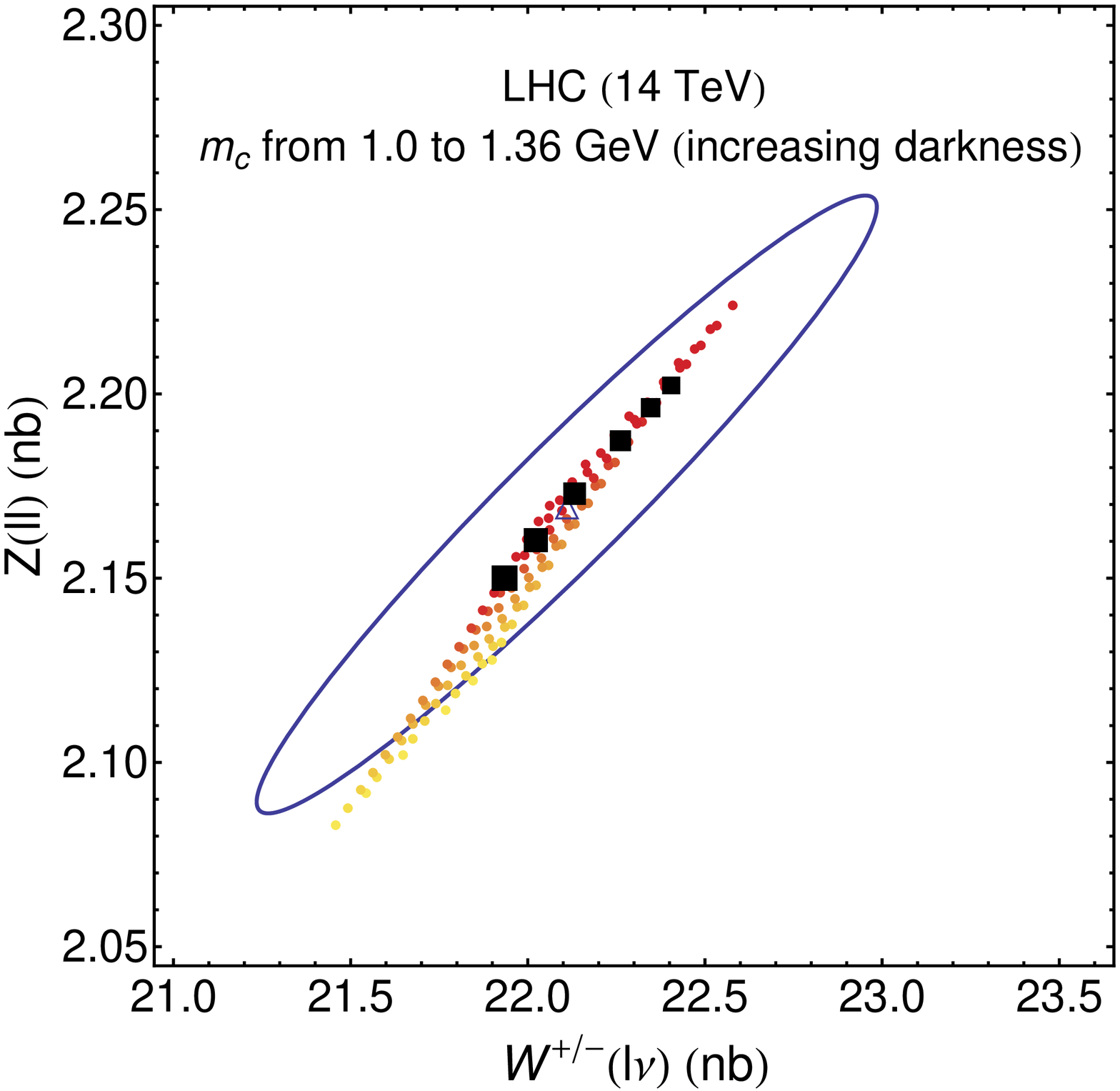}\\
\vspace{0.3in}
\includegraphics[width=0.395\textwidth]{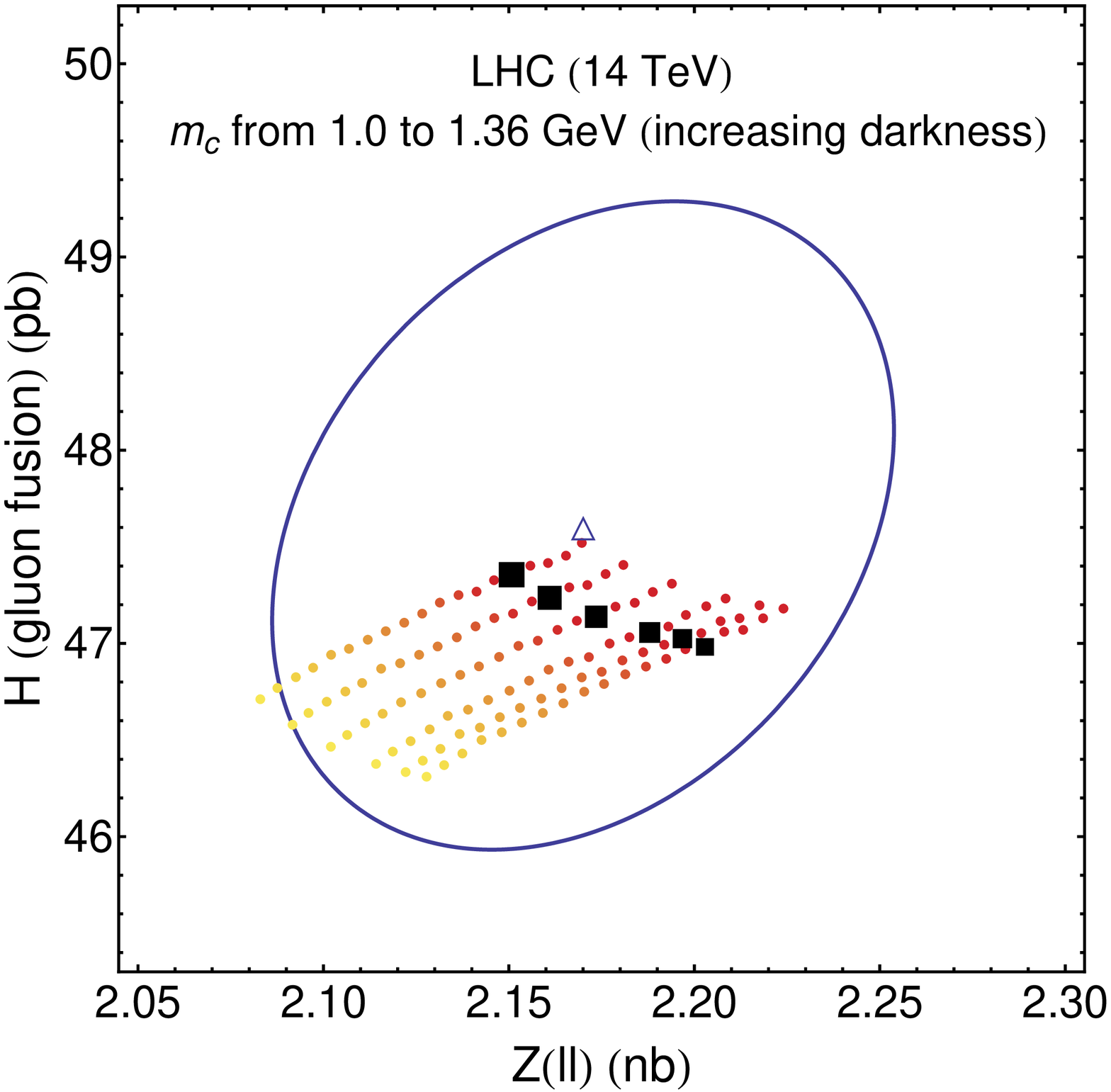} 
\includegraphics[width=0.41\textwidth]{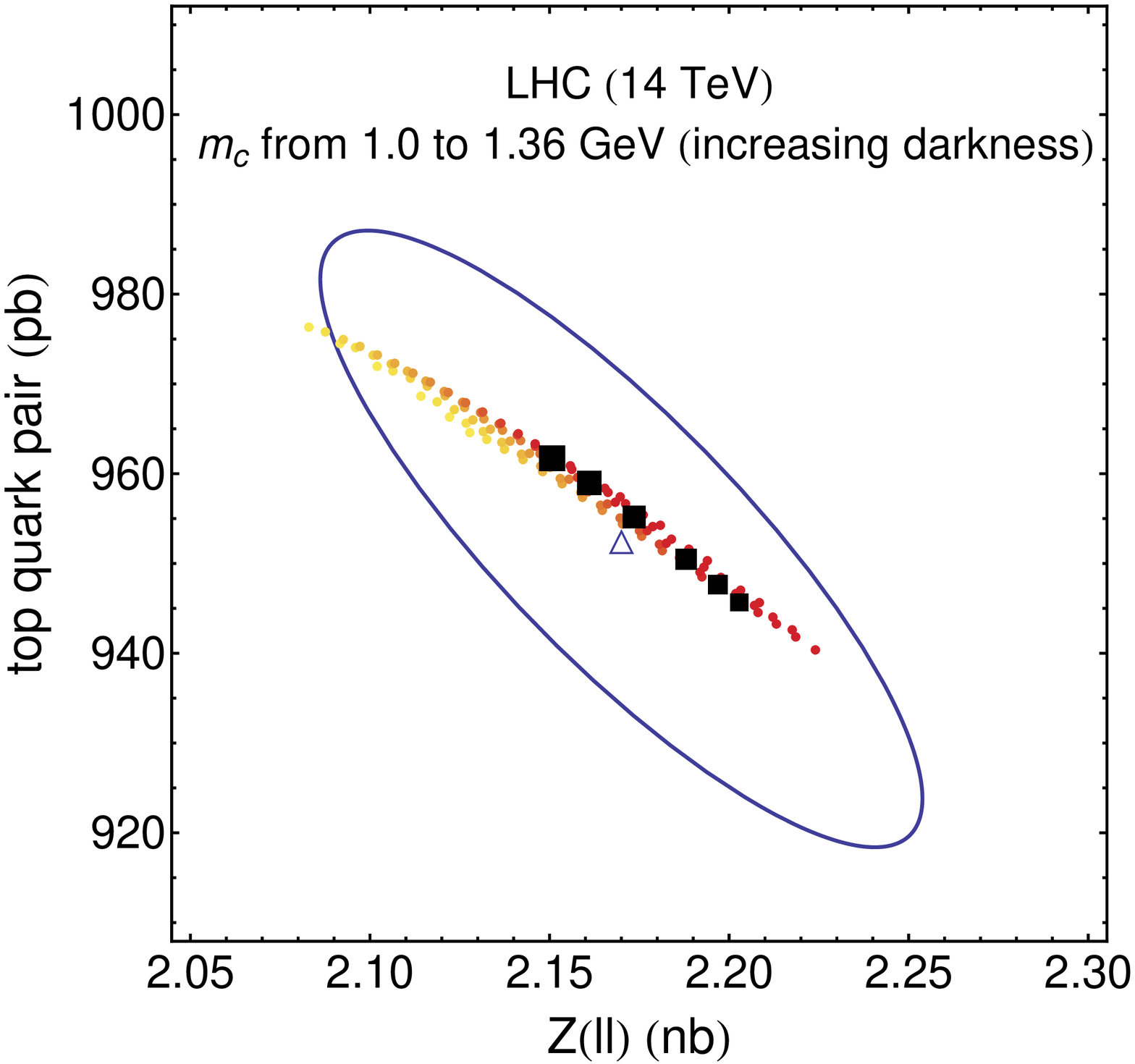} 
\par\end{centering}
\vspace{-1ex}
 \caption{\label{fig:wz2} Same as in Fig. \ref{fig:wz1}, but with $\sqrt{S}=14$ TeV.}

\end{figure}

PDF variations of such magnitude may have impact on collider observables
\cite{Nadolsky:2008zw}. For example, in Fig.~\ref{fig:wz1} and \ref{fig:wz2}
we show the dependence of NNLO total cross sections for $W,$ $Z,$ Higgs
boson production through gluon fusion, and top quark pair production at
the LHC at $\sqrt{S}=8$ and $14$ TeV. The NNLO cross sections for $W$ and $Z$
production are computed using FEWZ2.1~\cite{Gavin:2010az,Gavin:2012sy}.
The NNLO cross sections for
Higgs and top quark pair production are obtained from iHixs1.3~\cite{Anastasiou:2011pi} and
Top++2.0~\cite{Baernreuther:2012ws,Czakon:2013goa} with
$m_h=125\ {\rm GeV}$, $m_t=173.3\ {\rm GeV}$, and the QCD scales set to
the corresponding mass values. For each pair of total cross sections,
we show the central CT10NNLO prediction and an ellipse corresponding
to the 90\% C.L. PDF uncertainty interval of 
the CT10NNLO set~\cite{Gao:2013xoa}.
In the same figure, scattered points indicate the cross sections 
obtained with various combinations
of $m_{c}(m_{c})$ and $\lambda$ in the intervals $1\leq m_{c}(m_{c})\leq1.36$
GeV and $0\leq\lambda\leq0.2$. Here the darker colors
represent larger values of $m_{c}$, as in Fig.~\ref{fig:pdf}.
The $W$, $Z$, and Higgs production cross sections increase with
the charm mass by a few percent in the mass range considered. The trend is
opposite for $t\bar t$ production. The
changes are contained within the CT10NNLO PDF uncertainty ellipse,
however, and do not modify the PDF-induced (anti-)correlations observed between
the shown total cross sections.

The shown uncertainty in the LHC cross sections due to $m_{c}$ 
is comparable to the experimental PDF uncertainty and in principle
should be included independently from the latter. 
Let us outline one possibility for reducing
the $m_c$ uncertainty. Instead of allowing $m_{c}(m_c)$ to vary in the
whole interval $1 - 1.36$ GeV of its PDF uncertainty, 
we could set it to be at the world-average $m_{c}(m_{c})$
value of 1.275 GeV or in the $1\sigma$ interval $\pm0.025$ GeV around
it. This would reduce the associated uncertainty in the PDFs and
QCD observables. The corresponding predictions for the LHC $W,$ $Z,$
Higgs, and top quark pair production cross sections, obtained
for $m_{c}(m_{c})=1.28$ GeV and five explored $\lambda$ values, are shown in
Figs.~\ref{fig:wz1} and \ref{fig:wz2} by black boxes, 
with the size of the boxes increasing with the value of $\lambda$. 
The spread in these predictions constitutes only a part of 
the full span covered by the scattered points for the interval
$1\leq m_{c}\leq1.36$ GeV. It reflects only the uncertainty due to the
form of the rescaling variable, controlled by the $\lambda$
parameter. Theoretical predictions are better clustered in this case.

\section{Conclusions}  
We explored the charm quark mass dependence in the CTEQ NNLO global
PDF analysis that includes the recently published combined data on
charm quark production at the $ep$ collider HERA. This analysis, carried
out in the S-ACOT-$\chi$ heavy-quark scheme at order $\alpha_s^2$, 
renders an optimal $\overline{\rm MS}$
charm mass that is compatible with the world-average value $m_{c}(m_{c})=1.275\pm 0.025$
GeV. For example, with the $\overline{\rm MS}$ Wilson coefficient
functions for NC DIS, we obtain $m_c(m_c)=1.19^{+0.08}_{-0.15}
\mbox{ GeV}$, where the errors indicate the 68\% C.L. uncertainty due
to the PDFs and variation of the rescaling variable, as well as the
scale and $\alpha_s$ uncertainties added in quadrature. 

In QCD predictions for massive-quark DIS, 
one draws a distinction between the $\overline{\rm MS}$
charm mass, the fundamental parameter of the QCD Lagrangian, and auxiliary
energy scales of order of the physical charm mass. The auxiliary
scales that can contribute are specified by the factorization scheme. They
can be associated with 
the evolution of the QCD coupling strength, evolution of PDFs, heavy-quark
fragmentation, and powerlike contributions
$(m_{c}/Q)^{p}$ in DIS coefficient functions with initial-state charm
quarks. 

We argue that the sensitivity to the auxiliary scales is reduced
as the order of the PQCD calculation increases. In support of this
argument, Fig.~\ref{fig:chi2separate} demonstrates that the DIS
data are sensitive mostly to the physical mass
parameter in the exact DIS coefficient functions and less to the auxiliary
mass scales in the other parts of the calculation. 
Thus, the hadronic
cross sections in DIS and other processes at NNLO and beyond become
increasingly suitable for the determination of the fundamental charm mass.

The main findings of our fit are summarized in Fig.~\ref{fig:bench}
showing the 68\% C.L. intervals for the $\overline{\rm MS}$ charm mass 
obtained under various assumptions explained in the text. The central value
of $m_{c}(m_{c})$ in the S-ACOT-$\chi$ scheme at two loops tends to undershoot 
the world-average value according to the figure, but is compatible 
with the latter within the uncertainty. 

The uncertainties in the $m_{c}(m_{c})$ determination of 
both experimental and theoretical origin were explored 
in Sec.~\ref{sec:Results}. If $m_{c}(m_{c})$ 
is varied as an independent parameter 
in the full range of order 1-1.4 GeV 
allowed by the NNLO PDF fit, it increases the uncertainty
on the resulting PDFs, compared to a fixed $m_c$. 

For comparison, the
accuracy of the world-average $m_{c}(m_{c})$, at about $0.025$ GeV, 
is smaller than the NNLO fit uncertainty. 
By using a constant value of the $m_{c}(m_{c})$ parameter,
for example by setting it equal to its world-average value, 
one can suppress the
corresponding uncertainty in the PDFs. 
This strategy is similar to implementing the QCD coupling dependence 
in the global analysis \cite{Lai:2010nw}, 
when using the world-average value of $\alpha_{s}(M_{Z})=0.118$
results in tighter constraints on the PDFs than in a fit with a free
$\alpha_{s}(M_{Z})$. When the
input $m_{c}(m_{c})$ value is held constant instead of being fitted, 
one suppresses
the associated uncertainty in the PDFs and LHC cross sections, see
sample calculations in Figs.~\ref{fig:wz1} and
\ref{fig:wz2}. The residual
theoretical uncertainty for a fixed $m_{c}(m_{c})$ then arises only
from variations in the auxiliary scales and powerlike contributions
and can be suppressed by including higher orders in $\alpha_{s}$. 
\begin{acknowledgments}
This work was supported by the U.S. DOE Early Career Research Award
DE-SC0003870 by Lightner-Sams Foundation. 
We thank Achim Geiser for
the critical reading of the manuscript and appreciate 
detailed discussions with Karin Daum, 
Joey Huston, Hung-Liang Lai, Katerina Lipka, Fred Olness, 
Jon Pumplin,  Carl Schmidt, Dan Stump, and C.-P. Yuan. 
P.~N. thanks DESY
(Hamburg) for hospitality and financial support of his visit during 
the work on this project. 
\end{acknowledgments}

\bibliographystyle{apsrev}

\end{document}